\documentclass[reprint,amsmath,amssymb,aps]{revtex4-2}
\pdfoutput=1

\usepackage{graphicx}
\usepackage[usenames,dvipsnames]{xcolor}
\usepackage[colorlinks=true, linkcolor=black, citecolor=ForestGreen]{hyperref}
\usepackage{dcolumn}
\usepackage{bm}
\usepackage{overpic,subfigure}
\usepackage{cancel}

\definecolor{inkred}{RGB}{210,29,0}
\definecolor{inkblue}{RGB}{0,112,196}
\newcommand{\CO}{{\cal O}}
\newcommand{\<}{\langle}
\renewcommand{\>}{\rangle}

\makeatletter
\def\l@subsubsection#1#2{}
\makeatother

\begin{document}

\preprint{APS/123-QED}

\title{Thermalization and chaos in a 1+1d QFT}

\author{Luca V.~Delacr\'etaz}
	\email{lvd@uchicago.edu}
\affiliation{
Kadanoff Center for Theoretical Physics and Enrico Fermi Institute\\
University of Chicago, Chicago, IL 60637,
USA
}%

\author{A.~Liam Fitzpatrick}%
 \email{fitzpatr@bu.edu}
\author{Emanuel Katz}%
 \email{amikatz@bu.edu}
\affiliation{%
 Department of Physics, Boston University, Boston, MA 02215, USA
}%

\author{Matthew T.~Walters}%
 \email{matthew.walters@epfl.ch}
\affiliation{%
 Institute of Physics, \'Ecole Polytechnique F\'ed\'erale de Lausanne (EPFL)\\
 CH-1015, Lausanne, Switzerland\\
 Department of Theoretical Physics, Universit\'e de Gen\`eve\\
 CH-1211, Gen\`eve, Switzerland
}%

\date{\today}

\begin{abstract}
We study aspects of chaos and thermodynamics at strong coupling in a scalar model using LCT numerical methods.  We find that our eigenstate spectrum satisfies Wigner-Dyson statistics and that the coefficients describing eigenstates in our basis satisfy Random Matrix Theory (RMT) statistics.  At weak coupling, though the bulk of states satisfy RMT statistics, we find several scar states as well.  We then use these chaotic states to compute the equation of state of the model, obtaining results consistent with Conformal Field Theory (CFT) expectations at temperatures above the scale of relevant interactions. We also test the Eigenstate Thermalization Hypothesis by computing the expectation value of local operators in eigenstates, and check that their behavior is consistent with thermal CFT values at high temperatures.  Finally, we compute the Spectral Form Factor (SFF), which has the expected behavior associated with the equation of state at short times and chaos at long times.  We also propose a new technique for extracting the connected part of the SFF without the need of disorder averaging by using different symmetry sectors.
\end{abstract}

\maketitle

\tableofcontents

\section{Introduction and Synopsis}
The emergence of thermodynamics in closed quantum chaotic systems is a phenomenon which is still not fully understood.  For example, in what sense does an initial high energy state evolving with time begin to resemble a thermal bath?  The general belief is that correlation functions of local operators in such a state at sufficiently late times behave as they would at finite temperature.  The Eigenstate Thermalization Hypothesis (ETH) has become the guiding framework for parameterizing such thermalization \cite{PhysRevA.43.2046,PhysRevE.50.888,rigoleth}. 
In particular, it posits that a generic high energy eigenstate of the Hamiltonian is effectively already a thermal bath as far as local observables are concerned.  Thus, time dependent correlation functions in an eigenstate (or in a superposition of eigenstates in an energy band) should decay with time as expected at finite temperature, ultimately settling down to the equilibrium values predicted by the canonical ensemble.  Some of the above intuition has been tested numerically in quantum many-body systems on the lattice, however there have been far fewer tests of it in the context of QFTs.  The issue is finding reliable numerical methods for simulating the time evolution of excited QFT states.  Hamiltonian truncation offers the possibility of such a method, as it directly computes the eigenstates and their eigenvalues.  

Still, even with those eigenstates in hand, questions remain---both conceptual and practical.  Conceptually, it is not clear in light of chaos, what information any set of such eigenstates possesses, and how one should extract that information.  Specifically, truncation (or any other calculational scheme) will always introduce errors in the states, errors which are more significant than the typical splitting between the states (which is exponentially small in the volume and energy). Thus, one should always consider observables which either involve averaging over states, or are naturally self-averaging, in order to extract physically meaningful quantities. This leads to a further question of which kind of averaging procedure is optimal for a particular observable, and which observables are naturally more self-averaging. Relatedly, on a practical level, how does one establish in the chaotic regime---which is always non-perturbative---that the states one has computed are ``healthy''?

\begin{figure}
\centerline{
\begin{overpic}[width=0.8\linewidth,tics=10]{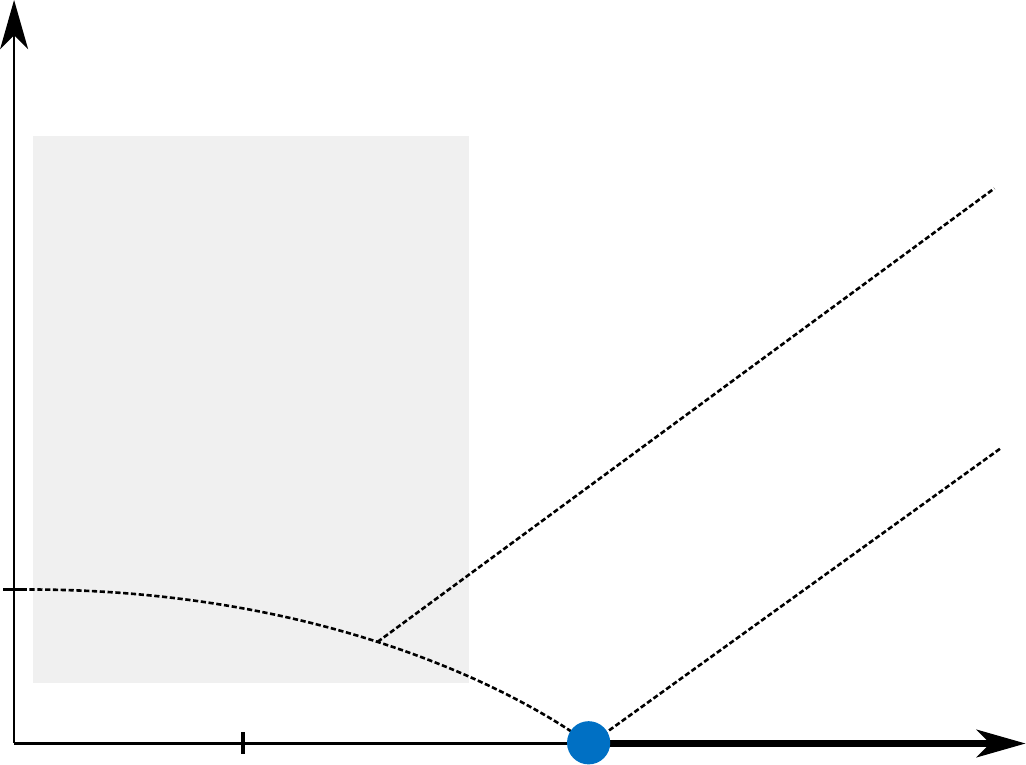}
	 \put (102,-0) {$\sqrt{\lambda}/m$} 
	 \put (-12,70) {$T/m$} 
	 \put (25,50) { \color{inkred}`QGP'} 
	 \put (43,12) {\color{inkblue} Ising QC fan} 
	 \put (75,-4) {$\cancel{\mathbb Z_2}$} 
	 \put (22,-4) {$1$} 
	 \put (11,7) {gapped} 
	 \put (-5,16) {$1$} 
\end{overpic}
}
\caption{\label{fig_phasediag} Thermal phase diagram of 1+1d $\lambda\phi^4$ theory. In the high temperature `QGP' region, the equilibrium thermodyanmics is approximately that of a free boson. Dashed lines are crossovers. This work explores the shaded parametric region.}
\end{figure}
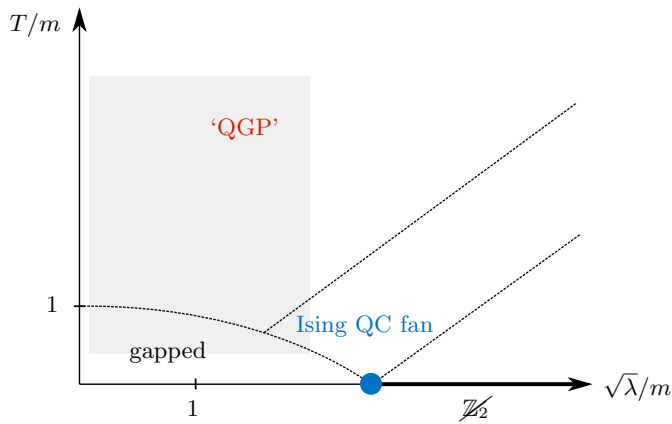

In this paper we address these questions and study highly excited states in the context of 1+1d $\phi^4$ theory:
\begin{equation}\label{eq_S}
S = \int d^2 x \, \frac12 (\partial\phi)^2 - \frac12 m^2 \phi^2 - \frac{1}{4!} 4\pi\lambda \phi^4\, .
\end{equation}
A schematic phase diagram expected for this theory is shown in Fig.~\ref{fig_phasediag}. This theory is asymptotically free, so that the high temperature equation of state $T\gg m,\, \sqrt{\lambda}$ is approximately that of a free scalar (this region is denoted `QGP' in a generous analogy with the quark-gluon plasma). At low temperatures, the equation of state is Boltzmann suppressed as the theory is gapped, except in the Ising quantum critical fan near the critical coupling where the degrees of freedom are those of a free fermion. The theory is expected to be most strongly thermalizing and chaotic away from these free limits; we will therefore be most interested in the regime $T\gtrsim \sqrt{\lambda}\gtrsim m \gg 1/V$.
The window in the figure indicates the region explored numerically in this work.  Thermal equilibrium and out-of-equilibrium properties of this model have been discussed in \cite{Delacretaz:2021ufg}, we review them in appendix~\ref{app_theory}.

To perform our computations, we use Lightcone Conformal Truncation (LCT), a Hamiltonian truncation method where one truncates on the scaling dimension, $\Delta$, of primary operators in a conformal basis \cite{Katz:2016hxp,Anand:2020gnn} (see appendix~\ref{app:LCTReview}).  Our approach is distinguished from most numerical work on quantum many-body dynamics in that it is directly formulated in the continuum rather than on the lattice, and continuous translation invariance is preserved exactly in our numerics. Indeed, in LCT, the basis states describe the free massless scalar CFT.  
Formally, these degrees of freedom live in infinite volume.  However, once we have added the relevant deformations (i.e.\ the mass and the quartic coupling), an effective volume emerges controlled by the truncation parameter $\Delta$.  We find that the effective volume of a state depends on its energy in a universal manner: $V\sim \Delta / E$, where $E$ is the rest frame energy of the state. 
In fact, in practical terms, it is advantageous that the volume changes throughout the spectrum, as it effectively allows LCT to access high temperature (or energy density) states at any truncation, as we will explain below.  Moreover, though there have been several Hamiltonian truncation approaches to thermalization in QFTs (see in particular Refs.~\cite{Brandino:2010sv,Rakovszky:2016ugs,Kukuljan:2018whw,Robinson:2018wbx,Srdinsek:2020bpq,Szasz-Schagrin:2021lal} for applications of these methods to out-of-equilibrium dynamics), the regime $T\gtrsim m,\,\sqrt{\lambda} \gg 1/V$, which we explore in this work has not been studied as far as we are aware.

Having understood the above volume dependence, we can then proceed to check the ``health'' of our states. First, we observe that our spectrum satisfies Wigner-Dyson statistics, indicating that the eigenstates are strongly coupled and chaotic (this is true even at weak coupling $\lambda/m^2 \ll 1$).  Second, using this chaotic spectrum, we compute the non-perturbative equation of state and various related observables for different choices of the coupling. These quantities exhibit Cardy behavior at high temperature, and at lower temperature agree with expectations for a CFT deformed by a relevant perturbation.  Next, we use the data contained in the eigenstate wavefunctions: we verify that the coefficients describing typical eigenstates in our basis satisfy Random Matrix Theory (RMT) statistics, and use those eigenstates to calculate the expectation value of local observables---we find that those values are consistent with those of a free scalar in a thermal bath, as predicted by ETH. These tests establish the thermality of individual eigenstates.  Note that this test of ETH, whereby we compare numerical expectation values to known analytic results for the free scalar thermal CFT, is possible because LCT allows us to access high temperature states.

We conclude by computing the simplest time-dependent observable which captures both features of the equation of state, as well as features of chaos---namely, the Spectral Form Factor (SFF), and the related survival probability.  Indeed, we find that it has a ``slope-ramp-plateau'' profile, with the slope determined by the equation of state and the ramp by RMT. We also introduce a `symmetry-resolved' SFF, which gives access to the connected SFF without the need of disorder averaging. 

Our numerical results for the model in Eq.~\eqref{eq_S} suggest that, taking symmetries properly into account, generic interacting QFTs thermalize and feature universal signatures of quantum chaos akin to quantum many-body systems on the lattice. At strong coupling, {\em all} high energy microstates are found to behave like thermal states (this can be contrasted to holographic QFTs where nonthermal or `scar' states exist \cite{Denef:2000nb,Anninos:2013mfa,Dodelson:2022eiz} despite strong thermalization and chaos). At weak coupling, we find instead a small collection of nonthermal (nonetheless non-perturbative) scar states.

Furthermore, we hope that this work is an initial step in a larger program to extend the reach of LCT methods towards the goal of understanding thermodynamics and non-equilibrium physics.  
In 1+1d, for any Lagrangian theory of scalars, fermions, and gauge fields, LCT methods allow one to calculate the spectrum.  Previously, analysis of the spectrum was concentrated on the low lying states.  However, as demonstrated in this work, one can use the high energy spectrum in order to determine the equation of state and related quantities.  In certain cases, for example, when the theory is chiral, supersymmetric, or contains CP-violating phases, Hamiltonian methods can have an advantage over lattice Monte-Carlo in determining the equation of state.  In addition to the spectrum, Hamiltonian methods also calculate the eigenfunctions of high energy states.  Averaging over these eigenfunctions one can begin to approximate the thermal expectation values of local operators.  Finally, one can attempt to use these eigenfunctions to calculate time dependent quantities, such as correlation functions in excited states.  Such correlation functions should have emergent hydrodynamic behavior that can teach us about thermalization and transport.  Though computationally more difficult, LCT technology is also available for calculating the above observables in 2+1d, where much less has been tested about chaos and thermalization.  Hence, there are many opportunities for progress in the future.

\section{Thermodynamics}\label{sec_thermo}

The basic idea of our approach is to use the non-perturbative energy eigenstates to create a thermal bath by directly performing microcanonical and macrocanonical sums.  Because the truncation in LCT restricts us to a specific set of states in the theory, we want to understand when the distribution of these states produces sensible thermodynamic quantities.  Ultimately, of course, we are interested in studying the time evolution of deviations around thermal ensembles, and so studying thermality is a necessary preliminary step.

\subsubsection*{Intrinsic Scan Over Volumes}

One key conceptual point that we have to take into account is that the truncation effectively limits the internal size of each state---roughly speaking, our truncation acts like an energy-dependent volume. This can be understood intuitively by the fact that basis states are states in the free scalar CFT which take the schematic form $\partial^{\ell_1}\phi \partial^{\ell_2}\phi  \cdots \partial^{\ell_n}\phi|0\rangle$, with number of derivatives and fields bounded by the truncation $\Delta$. One therefore expects these states (or operators) to have a spatial extent which scales with $\Delta$ \footnote{This can be seen heuristically by analogy with the lattice, where size is proportional to the number of discrete derivatives. The LCT basis in fact bears a certain resemblence with the Krylov space of operators that can be built on the lattice, also in the infinite volume limit, see for example Ref.~\cite{Parker:2018yvk}.}. In fact, in their Fock space description, these basis states are multi-particle states with wavefunctions that are polynomials in the momenta of the individual particles, and the order of the polynomials is given by the number of derivatives of the fields in the operator description \cite{Anand:2020gnn}.  The resolution in momentum space of such a basis is inversely proportional to the order of the polynomial, again indicating that the maximum spatial extent should scale with $\Delta$.  Dimensional analysis then suggests that the volume of an eigenstate $|E \rangle$ built out of such states scales as $V\sim \Delta/E$. By directly measuring eigenstates (see appendix \ref{app:SizeOfStates}), we find that they indeed have a physical size $V$ that smoothly depends on the energy of the state as
\begin{equation}\label{eq_volume}
V = \alpha \frac{\Delta}{E}, 
\end{equation}
where the dimensionless proportionality constant $\alpha$ depends on exactly how the size is defined, but has very little dependence on the state itself (for this reason, we will sometimes omit $\alpha$ in the following).  In order to focus on the thermal properties of states, we relegate a detailed discussion of how we extract the size $V$ of states to appendix \ref{app:SizeOfStates}.

An interesting feature of the volume-dependence (\ref{eq_volume}) of our states is that at a fixed truncation $\Delta$, the volume is intrinsically scanned over by the states themselves.  As a consequence, one automatically sees states with large temperatures, for any value of the truncation parameter.  Because the volumes are determined partially by dynamics rather than by hand, one might worry that most of the time they could end up being too small, but in fact for almost all states the volumes are larger than the inverse temperature.  The reason is simply that $V T \sim \sqrt{E V}$ is a nearly energy-independent constant, given by the total entropy $S \sim \sqrt{\Delta} \gg 1$.  Moreover, the fact that the entropy $S$ or density of states $e^S$ is energy independent at fixed $\Delta$ usefully implies that the truncation produces roughly an equal number of states per decade in energy, so that with a single diagonalization of the Hamiltonian at fixed $\Delta$ one can study a large range in temperatures.  By contrast, more traditional truncations at fixed volume produce states whose density is heavily weighted toward the highest energy scales in the truncation, since $S(E) \sim \sqrt{E}$.

In addition to $V\gg 1/T$, one must also ensure that $V\gtrsim 1/m$ to study the QFT in the thermodynamic limit (the opposite regime $V<1/m,\,1/\sqrt{\lambda}$ instead probes the UV CFT, with very small deformation leading to a { local} equilibration time larger than the volume). In terms of the energy density $\varepsilon = E/V \sim E^2/\Delta$, this implies that we expect to find thermal states in the window $1\lesssim \frac{\varepsilon}{m^2} \lesssim {\Delta}$. The loose lower bound comes from the fact that Boltzmann suppression will make low temperature states thermalize inefficiently. In practice, we will find that the window of strongly thermalizing states is somewhat larger.

Knowing the size of states allows us to convert extensive quantities, such as energy and entropy in particular, into densities, by dividing by the volume $V$.  Moreover, the energy-dependence of the volume of states must be taken into account when we extract standard thermodynamic quantities that are defined at fixed volume.  In particular,
\begin{equation}
T dS = dE + P dV = \left( 1 + P \frac{\partial V}{\partial E} \right) dE + P \frac{\partial V}{\partial \Delta} d \Delta .
\label{eq:thermDeltaMax1}
\end{equation}

\subsubsection*{Microcanonical Approach}

\begin{figure}
\centerline{
\begin{overpic}[width=0.9\linewidth,tics=10]{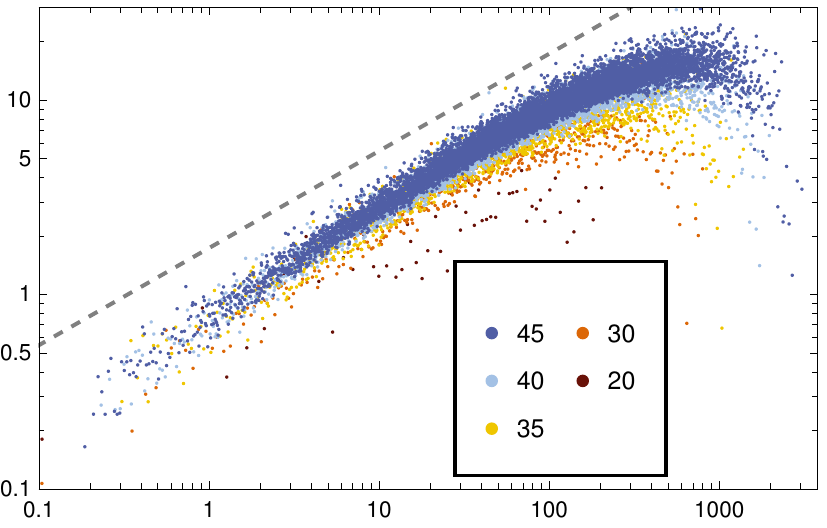}
	 \put (-5,30) {\rotatebox{90}{$s(\varepsilon)/m$}} 
	 \put (52,-2) {$\varepsilon/m^2$} 
	 \put (66.5,26) {$\Delta$} 
\end{overpic}
}
\caption{\label{fig_dos} Cardy-like growth of the density of states, as measured by the entropy density $s$. The dashed gray line $\sim \sqrt{\varepsilon}$ is a guide to the eye. The density of states ceases to grow at high energies due to truncation of the Hilbert space. ($\lambda = 1.6m^2$). 
}
\end{figure}

One way to extract the entropy $S$ at a given energy is to simply take the log of one over the density of states $\rho(E)$, which is straightforward to calculate given the set of energy eigenvalues $E_i$.  More precisely, we can define $\frac{1}{\rho(E)} = E_{i+1} - E_i$, where $E_i < E < E_{i+1}$.  Then, for each state, we obtain an  entropy density $s\equiv S/V$ and energy density $\varepsilon \equiv E/V$, which we plot in Fig.~\ref{fig_dos} for a specific choice of coupling $\lambda/m^2 = 1.6$.  We find that, over multiple decades in $\varepsilon$, the entropy density exhibits Cardy behavior:
\begin{equation}\label{eq_Cardy}
s(\varepsilon) \sim \sqrt{\varepsilon}.
\end{equation}
We see deviations from this behavior at small and large values of $\varepsilon$.  Deviations at small $\varepsilon/m^2$ are expected even in the absence of truncation, because of the physical mass gap, whereas deviations at large $\varepsilon/m^2$ are due to the fact that at high enough energies one simply runs out of states in the truncated spectrum.  In fact, as one can see from Fig.~\ref{fig_dos}, the Cardy-like growth of states extends to higher energies as the truncation parameter $\Delta$ increases.

This microcanonical approach produces the entropy $S(E,\Delta)$ as a function of energy $E$ and $\Delta$, from which one can extract the temperature. By inspection of  (\ref{eq:thermDeltaMax1}),
\begin{equation}\label{eq_thermDelta}
\left(\frac{ \partial S}{\partial E} \right)_\Delta = \beta \left(1 - \frac{\alpha P \Delta}{E^2}\right), \quad \left(\frac{ \partial S}{\partial \Delta} \right)_E= \beta\frac{\alpha P}{E},
\end{equation}
 using $V = \alpha \Delta/E$.  

Therefore, by computing how the density of states varies with $\Delta$ and $E$ separately, one can 
extract a pressure $P$ and temperature $T$ for each state. However, in this paper we will not pursue this microcanonical equation of state any further.  Empirically, we have found that it converges somewhat slowly with increasing $\Delta$, and in practice a more efficient method for obtaining thermodynamic quantities is to construct canonical ensembles.

\subsubsection*{Canonical Approach}

In a canonical approach, we first construct a partition function $Z$ for our spectrum obtained at a fixed truncation $\Delta$:
\begin{equation}
Z(\tilde{\beta}, \Delta) \equiv \sum_n e^{- \tilde{\beta} E_n} \simeq \int dE \exp \left[ S(E, V(E,\Delta)) - \tilde{\beta} E\right],
\end{equation}
where the approximation of $Z$ as an integral holds when the spectrum is sufficiently dense. In the saddle point approximation, the integral is dominated by energies $E = E_*(\beta)$ where 
\begin{equation}
\tilde{\beta}  =  \frac{d S(E, V(E,\Delta))}{dE} \equiv  \left( \frac{\partial S}{\partial E} \right)_\Delta  .
\end{equation}
The parameter $\tilde{\beta}$ can be related to the usual inverse temperature $\beta$ by comparing with (\ref{eq_thermDelta}):
\begin{equation}
\tilde{\beta} = \beta \left( 1 - \frac{\alpha P \Delta}{E^2} \right)= \beta \left( 1 - \frac{P}{\varepsilon} \right),
\label{eq:BetaTildeVsBeta}
\end{equation}
where $\varepsilon = E/V$. Moreover, $P$ and $E$ can be obtained from $\log Z(\tilde{\beta}, \Delta) \sim S(E,V(E,\Delta)) - \tilde{\beta} E$ through
\begin{equation}
 \partial_{\tilde{\beta}} \log Z(\tilde{\beta}, \Delta) = -E, \quad \partial_\Delta \log Z(\tilde{\beta}, \Delta) = \frac{\alpha \beta P }{E}.
 \label{eq:CanonicalTherms}
\end{equation}
Equations (\ref{eq:BetaTildeVsBeta}) and (\ref{eq:CanonicalTherms}) are three equations that can be used to determine the three unknowns $P,E$ and $\beta$.  In particular,
\begin{equation}
\beta = \tilde{\beta} - \Delta \frac{\partial_\Delta \log Z}{\partial_{\tilde{\beta}} \log Z}, \  \frac{P}{\varepsilon} = 1 - \frac{\tilde{\beta}}{\beta},  \textrm{ and  } \ \frac{s}{\varepsilon} = \tilde{\beta} - \frac{\log Z}{\partial_{\tilde{\beta}} \log Z}
\end{equation}
are independent of the coefficient $\alpha$ in the volume.




It is useful to consider the dimensionless entropy density $s_o(T) \equiv s(T)/T$, which was argued in \cite{Delacretaz:2021ufg} to be a thermodynamic $C$-function in 1+1d QFTs; that is, it is monotonically increasing between fixed points, and at fixed points it takes the constant value $\pi c/3$ in terms of the central charge $c$. In the present context, there is an $O(1)$ unknown prefactor $\alpha$ in the volume of states \eqref{eq_volume}, and therefore also in $s_o$, so only the overall shape of $s_o(T)$ is meaningful.

By contrast, we can instead consider volume-insensitive observables, for which this $O(1)$ ambiguity factors out.  Two natural such quantities are the dimensionless pressure-to-energy density ratio and the speed of sound $c_s$.  In order to extract $c_s^2$, we need to take derivatives with respect to $\varepsilon$ at fixed $V$.  Because $V^2 = \frac{\alpha \Delta}{\varepsilon}$, along surfaces of fixed $V$ we have $d \Delta = \frac{\Delta}{\varepsilon} d \varepsilon$ and $(\partial_{\tilde{\beta}} \varepsilon )\  d \tilde{\beta} =(1 -\frac{\Delta}{\varepsilon} \partial_\Delta \varepsilon) d \varepsilon$.  It follows that
\begin{equation}
c_s^2(\tilde{\beta}, \Delta) \equiv \left( \frac{\partial P}{\partial \varepsilon}\right)_V  = \frac{1- \frac{\Delta}{\varepsilon}\partial_\Delta \varepsilon }{\partial_{\tilde{\beta}} \varepsilon} \partial_{\tilde{\beta}} P + \frac{\Delta}{\varepsilon} \partial_\Delta P.
\end{equation}

\begin{figure}
\centerline{
\subfigure{
\begin{overpic}[width=0.9\linewidth,tics=10]{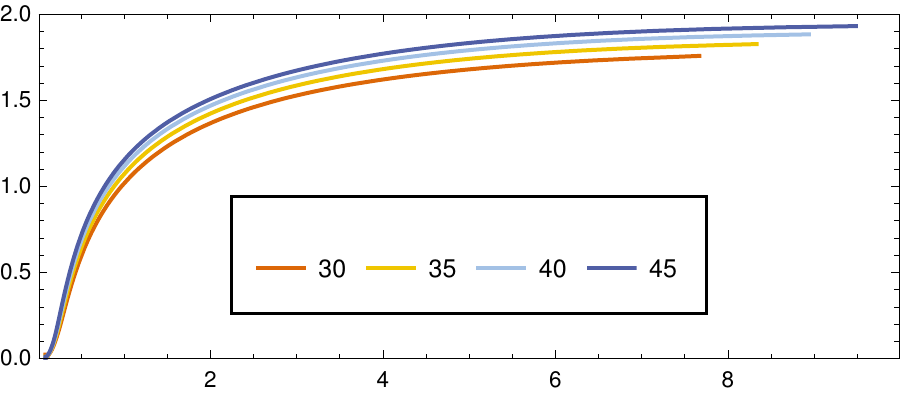}
	 \put (-6,23) {\rotatebox{90}{$s_o$}} 
	 \put (48,17.5) {$\Delta$} 
\end{overpic}
}}\vspace{-18.5pt}
\centerline{
\subfigure{
\begin{overpic}[width=0.9\linewidth,tics=10]{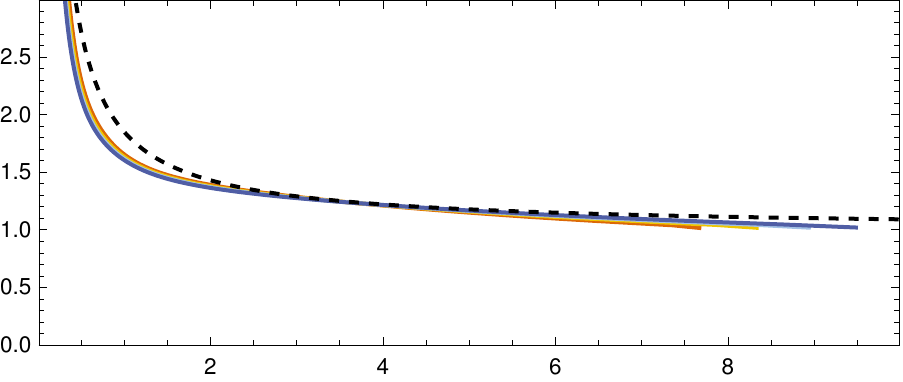}
	 \put (-6,18) {\rotatebox{90}{$\varepsilon/P$}} 
\end{overpic}
}}\vspace{-18.5pt}
\centerline{
\subfigure{
\begin{overpic}[width=0.9\linewidth,tics=10]{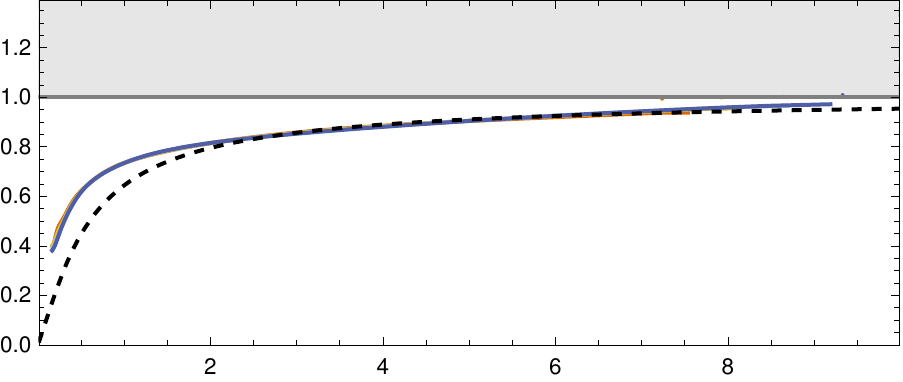}
	 \put (-6,23) {\rotatebox{90}{$c_s^2$}} 
	 \put (45,-3) {$T/m$} 
\end{overpic}
}}
\caption{\label{fig_eos} Equation of state in the interacting theory \eqref{eq_S} with coupling $\lambda=1.6m^2$. 
The dashed line shows the equation of state of a free massive scalar ($\lambda=0$) in the thermodynamic limit, for comparison; the interacting case is expected to deviate from this prediction when $T\lesssim \sqrt{\lambda}$.
}
\end{figure}

In Fig.~\ref{fig_eos}, we show $s_o, \frac{\varepsilon}{P}$, and $c_s^2$, as a function of $T$ at strong coupling $\lambda/m^2 = 1.6$ for various truncations.  The dimensionless entropy density $s_o$ still has a weak but visible dependence on $\Delta$ at $\Delta=45$, whereas the latter two observables appear to have converged fairly well even by $\Delta=30$. At high temperatures $T\gg m,\,\sqrt{\lambda}$, these approach their CFT values $\frac{\varepsilon}{P} = c_s^2 = 1$, and receive corrections from the relevant deformations at lower temperatures.  Notice that the speed of sound is subluminal in the temperatures of interest, consistent with causality. While our numerical approach preserves exact boost invariance, it is not manifestly local and the speed of sound obtained from the spectrum is not automatically smaller than one. Observing $c_s^2\leq 1$ therefore serves as a useful consistency check. At intermediate temperatures, the measured speed of sound is close to the one expected for a free massive scalar field (shown in dashed in Fig.~\ref{fig_eos}). At lower temperatures $T/m\lesssim 2$, $c_s^2$ exceeds its value for a free massless scalar, as expected: the interaction $\lambda = 1.6m^2$ reduces the mass gap of the model, leading to less Boltzmann suppression at low temperatures. In principle, if the coupling $\lambda$ is increased to its critical value where the model realizes the 1+1d Ising CFT, the speed of sound should in fact feature an upturn at low temperatures and return to the CFT value $c_s^2 = 1$ for $T\to 0$. While many aspects of the Ising IR CFT can be captured with LCT \cite{Anand:2017yij,Anand:2020gnn,Chen:2021pgx}, we do not have enough resolution to observe the Cardy density of Ising states at low energies, which is responsible for this upturn.

\subsubsection*{Expectation Values}

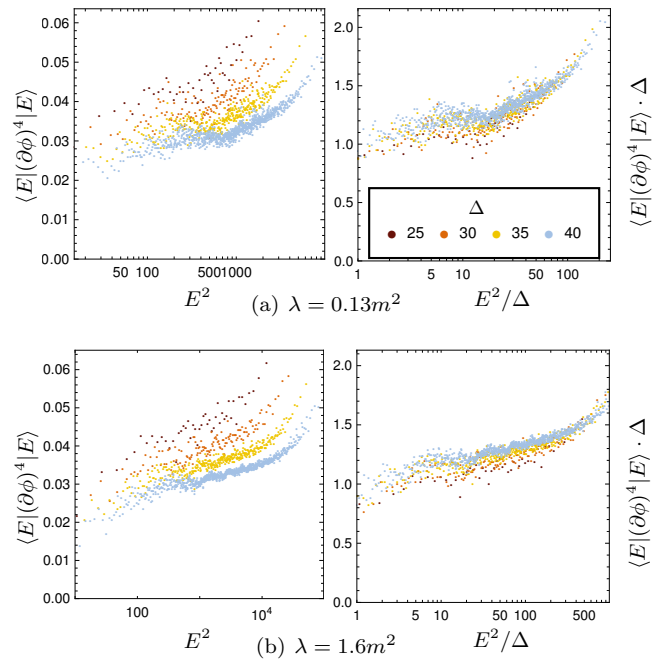
\begin{figure}
\centerline{
\subfigure[$\lambda=0.13m^2$]{
\begin{overpic}[height=0.43\linewidth,tics=10]{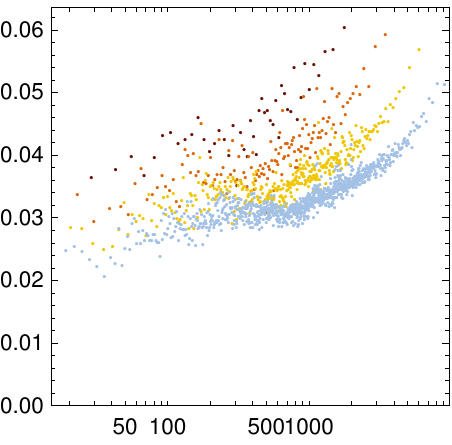}
	 \put (-10,28) {\scalebox{0.9}{\rotatebox{90}{$\langle E|(\partial\phi)^4|E\rangle$}}} 
	 \put (50,-8) {\scalebox{0.9}{$E^2$}} 
\end{overpic}
\begin{overpic}[height=0.43\linewidth,tics=10]{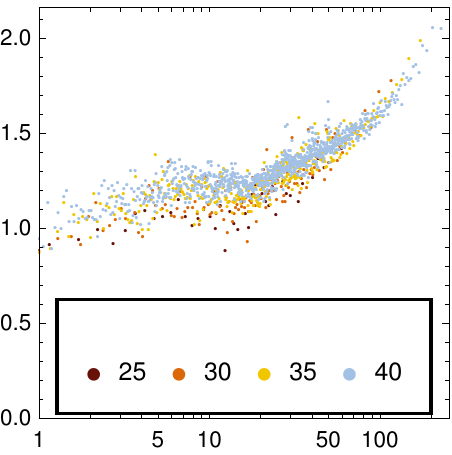}
	 \put (48,25) {\scalebox{0.8}{$\Delta$}} 
	 \put (50,-8) {\scalebox{0.9}{$E^2/\Delta$}} 
	 \put (105,20) {\scalebox{0.9}{\rotatebox{90}{$\langle E|(\partial\phi)^4|E\rangle \cdot \Delta$}}} 
\end{overpic}
}}
\centerline{
\subfigure[$\lambda=1.6m^2$]{
\begin{overpic}[height=0.43\linewidth,tics=10]{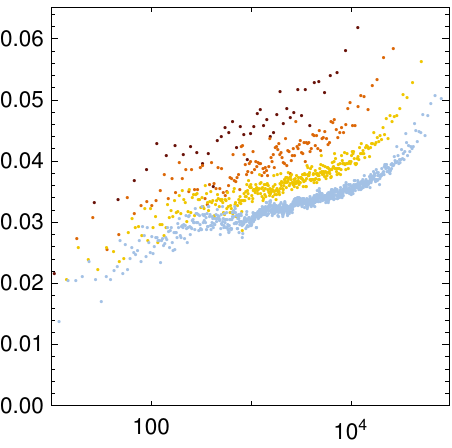}
	 \put (-10,28) {\scalebox{0.9}{\rotatebox{90}{$\langle E|(\partial\phi)^4|E\rangle$}}} 
	 \put (50,-8) {\scalebox{0.9}{$E^2$}} 
\end{overpic}
\begin{overpic}[height=0.43\linewidth,tics=10]{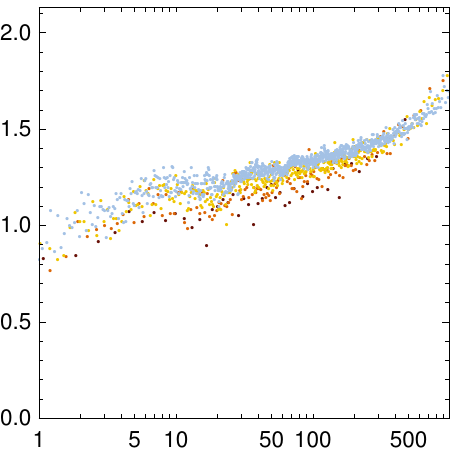}
	 \put (50,-8) {\scalebox{0.9}{$E^2/\Delta$}}
	 \put (105,20) {\scalebox{0.9}{\rotatebox{90}{$\langle E|(\partial\phi)^4|E\rangle \cdot \Delta$}}} 
\end{overpic}
}}
\caption{\label{fig_collapse} 
Scaling collapse of $(\partial\phi)^4$ expectation value at strong and weak coupling, when plotted against $E^2/\Delta$, as predicted by Eq.~\eqref{eq_collapse}. Each point is obtained from a microcanonical average over 20 states.
}
\end{figure}

In addition to these standard thermodynamic quantities,  the local operators of the theory  provide a much larger set of probes of the system.  In general, we expect that local operators in excited states should look approximately thermal, and in particular their expectation values should be given by a universal function of the energy density of the state.  Because LCT diagonalizes the Hamiltonian in a boosted frame, we must also account for the fact that operators with spin transform nontrivially when we boost back to the rest frame.  If, for an operator $\CO^{(J)}$ with spin $J$, we define the expectation value
\begin{equation}
\sum_{|E'-E|< \delta} \< E'  | \CO^{(J)} | E'\> = f_\CO(E, \Delta),
\label{eq:OpVevs}
\end{equation}
where $\delta$ sets the size of a small window in energies, then taking into account the boost to the rest frame (see appendix \ref{app:SizeOfStates}),  locality of $\CO^{(J)}$ suggests we can write $f_{\CO}(E,\Delta)$ as a function of energy density as
\begin{equation}\label{eq_collapse}
f_{\CO}(E, \Delta) = E^{J+2} f_{\CO}(E^2/\Delta) = ( \Delta)^{\frac{J+2}{2}} \tilde{f}_{\CO}(\varepsilon).
\end{equation}
Here, $\tilde{f}_\CO$ is defined so that we expect, up to numeric prefactors, that $\varepsilon^{-J/2}\tilde{f}_\CO(\varepsilon)$ is a thermal expectation value depending on $\Delta$ and $E$ only through the combination $\varepsilon$.  
In Fig.~\ref{fig_collapse}, we demonstrate that this expected scaling collapse holds at strong and weak coupling for the $J=-4$ operator $(\partial_- \phi)^4$. This collapse further confirms that the volume \eqref{eq_volume} has been correctly identified. Moreover, the small variance in these expectation values indicates that the ETH is satisfied for diagonal matrix elements. 

\begin{figure}
\centerline{
\begin{overpic}[width=0.85\linewidth,tics=10,trim={-0.12cm 0cm 0cm 0cm},clip]{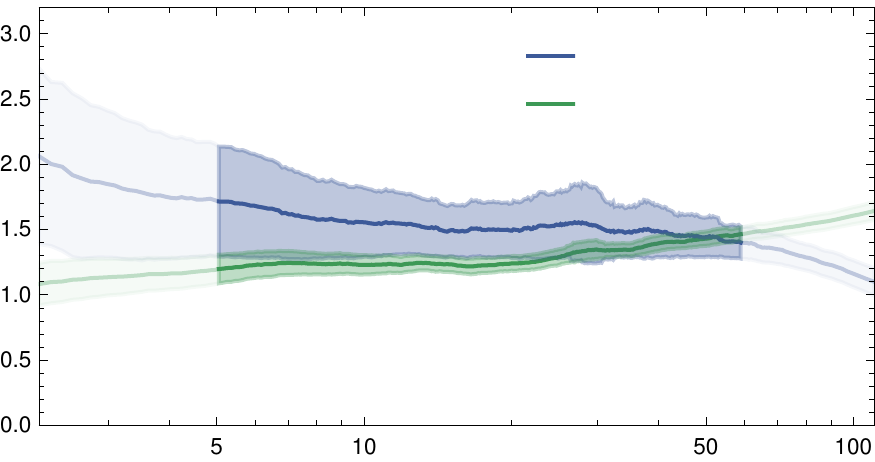}
	 \put (-7,25) {\rotatebox{90}{$\langle \mathcal{O}\rangle/\varepsilon^2$}} 
	 \put (69,45.5) {$\mathcal{O}_1 =  \frac{5}{8\pi}(\partial^2\phi)^2$} 
	 \put (69,39) {$\mathcal{O}_2 =(\partial\phi)^4$} 
\end{overpic}
}
\vspace{-10pt}
\centerline{
\begin{overpic}[width=0.85\linewidth,tics=10,trim={0.03cm 0cm 0cm 0cm},clip]{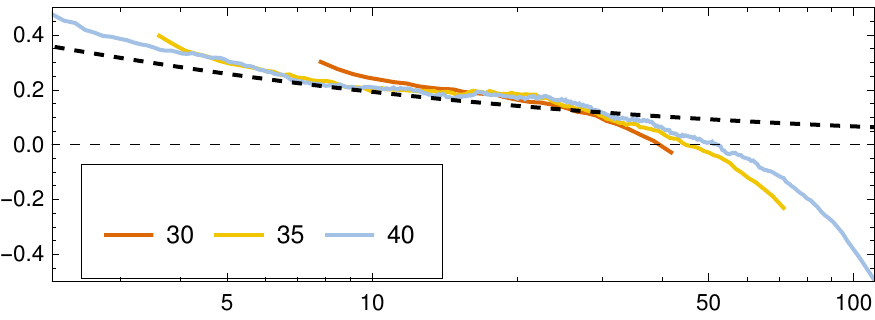}
	 \put (-6,5) {\rotatebox{90}{$\langle \mathcal{O}_1-\mathcal{O}_2\rangle/\langle\mathcal{O}_1\rangle$}} 
	 \put (25,13.5) {\scalebox{0.9}{$\Delta$}} 
	 \put (50,-3) {$\varepsilon/m^2$} 
\end{overpic}
}
\caption{\label{fig_WeakVirasoro} 
Top: Microcanonical expectation values of operators $\mathcal{O}_1=\frac{5}{8\pi}(\partial^2\phi)^2$ and $\mathcal{O}_2=(\partial\phi)^4$ at weak coupling $\lambda=0.13m^2$, with mean and standard deviation shown ($\Delta=40$). Both are seen to be $\propto \varepsilon^2$ and approximately equal for $\varepsilon/m^2 \gg 1$, as expected for the massless thermal scalar. The upper limit of the shaded region is where truncation effects become important. Bottom: a difference between the two expectation values is expected due to breaking of conformal invariance by the mass $m^2$ (and, to a lesser extent at weak coupling, $\lambda$). This difference agrees well with the analytic prediction for a free massive scalar ($\lambda=0$, dashed). 
}
\end{figure}

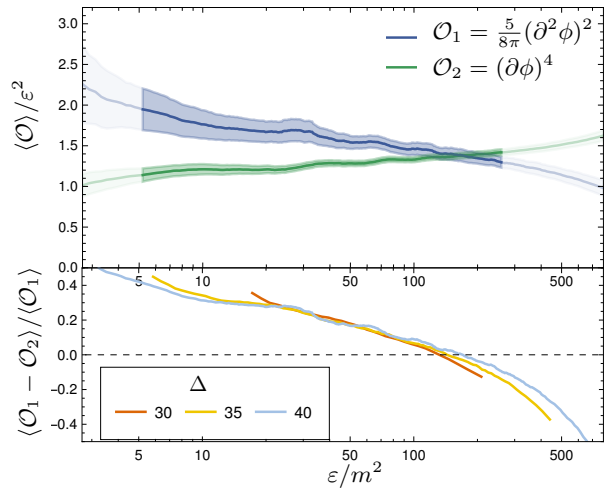
\begin{figure}
\centerline{
\begin{overpic}[width=0.85\linewidth,tics=10,trim={-0.12cm 0cm 0cm 0cm},clip]{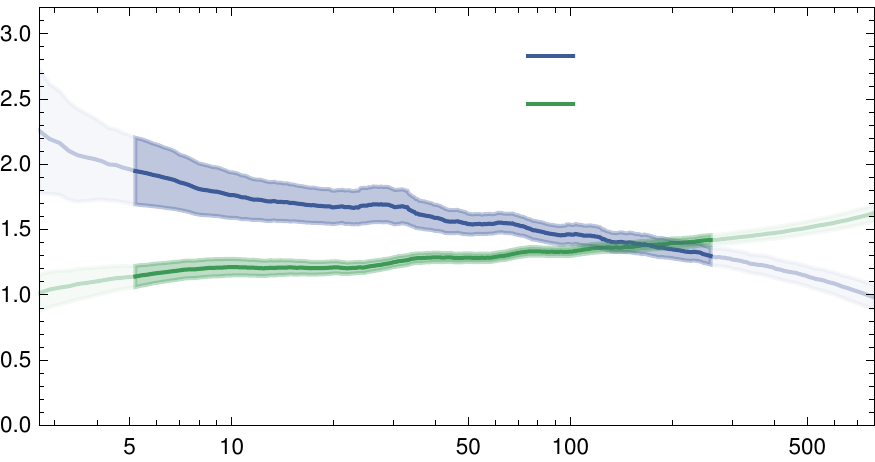}
	 \put (-7,25) {\rotatebox{90}{$\langle \mathcal{O}\rangle/\varepsilon^2$}} 
	 \put (69,45.5) {$\mathcal{O}_1 =  \frac{5}{8\pi}(\partial^2\phi)^2$} 
	 \put (69,39) {$\mathcal{O}_2 =(\partial\phi)^4$} 
\end{overpic}
}
\vspace{-10pt}
\centerline{
\begin{overpic}[width=0.85\linewidth,tics=10,trim={0.03cm 0cm 0cm 0cm},clip]{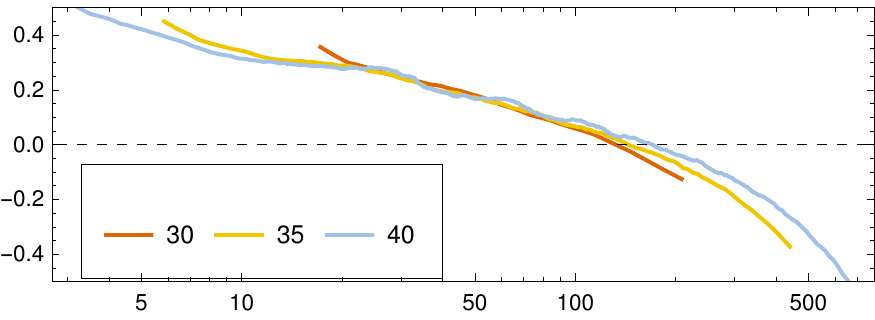}
	 \put (-6,5) {\rotatebox{90}{$\langle \mathcal{O}_1-\mathcal{O}_2\rangle/\langle\mathcal{O}_1\rangle$}} 
	 \put (25,13.5) {\scalebox{0.9}{$\Delta$}} 
	 \put (50,-3) {$\varepsilon/m^2$} 
\end{overpic}
}
\caption{\label{fig_Virasoro} 
Strong coupling version of Fig.~\ref{fig_Virasoro} ($\lambda=1.6 m^2$, $\Delta=40$). 
}
\end{figure}

One can compare these matrix elements to the theory prediction for the thermal expectation value. At high temperatures $T\gg m,\,\sqrt{\lambda}$ these are fixed to leading order by the thermal expectation values in the free scalar CFT. $(\partial_-\phi)^4$ is a Virasoro descendant of the identity operator and therefore can acquire a thermal expectation value. Since it is dimension 4, one has $\langle (\partial_-\phi)^4\rangle \sim T^4 \sim \varepsilon^2$. Figs.~\ref{fig_WeakVirasoro} and \ref{fig_Virasoro} show that this temperature dependence is indeed observed, in the same energy density region where where approximate Cardy growth was observed in Figs.~\ref{fig_dos} and \ref{fig_eos}. Finally, we would like to check the coefficient of the $\varepsilon^2$ dependence. Because the energy density is only determined up to the $O(1)$ factor in the volume in Eq.~\eqref{eq_volume}, we cannot compute that numerical coefficient directly; however the same ambiguity arises in the expectation value of any other operator and therefore drops out when comparing two expectation values. Consider for example another dimension-4 operator, $(\partial_-^2\phi)^2$. One can show (see app.~\ref{app_theory})
\begin{equation}
\langle (\partial_- \phi)^4\rangle
	= 3 \varepsilon^2\, , \qquad
\langle (\partial_-^2 \phi)^2\rangle
	= \frac{24\pi}{5}\varepsilon^2\, ,
\end{equation}
so that the thermal expectation value of the linear combination $(\partial_- \phi)^4 - \frac{5}{8\pi} (\partial_-^2 \phi)^2$ vanishes at high temperatures (this can be understood from the fact that this linear combination is a Virasoro primary, up to a total derivative). Figs.~\ref{fig_WeakVirasoro} and \ref{fig_Virasoro} show that the expectation value of this linear combination indeed approximately vanishes at high temperature, at weak and strong coupling. A deviation at lower temperatures is expected due to the breaking of conformal invariance by the dimensionful couplings $m^2$ and $\lambda$. At weak coupling, we find that the deviation indeed agrees with the one predicted for a free massive scalar (dashed line in Fig.~\ref{fig_WeakVirasoro}).  


Despite many numerical tests of ETH in local quantum-many body systems in the literature, this is as far as we know the first quantitative test against a thermal expectation value predicted from theory.

In appendix \ref{app_theory}, we also discuss the expectation value of the operator $\phi^2$, which has unusual properties due to the fact that in the free massive theory it is the trace of the stress tensor.

\section{Eigenvalue and Eigenvector Statistics}\label{sec_stats}

An important signature of chaotic systems is that the distribution of eigenvalues and eigenstates within small energy bands should be well-described by RMT statistics. For the specific case of 1+1d $\phi^4$ theory, whose Hamiltonian is time-reversal invariant, we specifically expect the eigenstates to be described by the Gaussian orthogonal ensemble (GOE).

\subsubsection*{Eigenvalue Spacings}

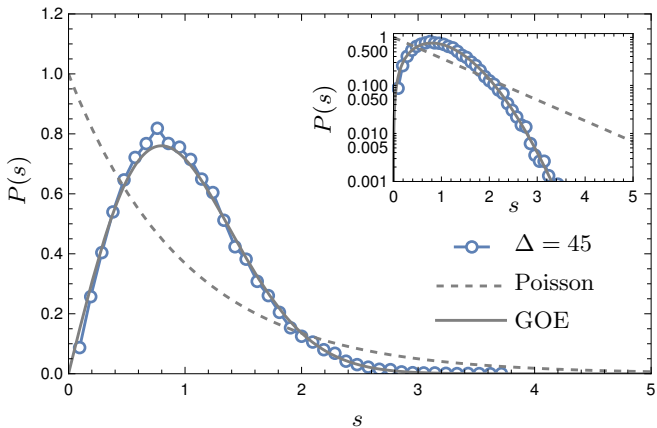
\begin{figure}
\centerline{
\begin{overpic}[width=\linewidth,tics=10,trim={0cm 0.2cm 0cm 0cm},clip]{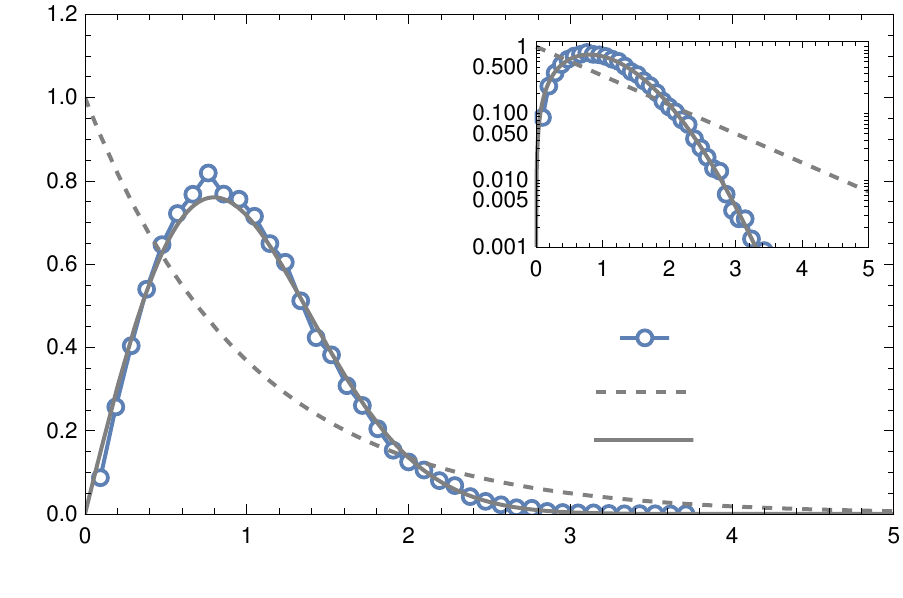}
	 \put (0,33) {\rotatebox{90}{$P(s)$}} 
	 \put (53,-1) {$s$} 
	 \put (47,44) {\small \rotatebox{90}{$P(s)$}} 
	 \put (77,32) {\small $s$} 
	 \put (78,26) {$\Delta=45$} 
	 \put (78,20) {Poisson} 
	 \put (78,14) {GOE} 
\end{overpic}
} 
\caption{\label{fig_WD} Statistics of eigenvalue spacing compared to Poisson and GOE Wigner-Dyson distributions, at strong coupling $\lambda=1.6m^2$ after unfolding the spectrum. Only states with energy density $1<\varepsilon/m^2<100$ were kept. 
The inset shows the same data on a logarithmic scale.
}
\end{figure}

Perhaps most famously, if one considers the spacing of adjacent eigenvalues in a given energy range (normalized by the average space $\overline{\delta E}$ in that range, i.e.~``unfolded''~\cite{mehta2004random}):
\begin{equation}\label{eq_unfolded}
s_i \equiv \frac{E_{i+1} - E_i}{\overline{\delta E}},
\end{equation}
the expectation is that if this theory is chaotic the distribution of these spacings should follow the GOE Wigner-Dyson distribution
\begin{equation}
P_{\textrm{GOE}}(s) = \frac{\pi}{2} s e^{-\frac{\pi}{4} s^2}.
\end{equation}

Fig.~\ref{fig_WD} shows the distribution of unfolded eigenvalue spacings for 1+1d $\phi^4$ theory at strong coupling ($\lambda=1.6m^2$) obtained with LCT at $\Delta=45$. In order to focus specifically on the ``healthy'' thermal states satisfying Cardy growth in Fig.~\ref{fig_eos}, we have restricted this distribution to eigenstates with $1 < \varepsilon/m^2 < 100$. As we can see, the truncation data (blue circles) clearly follows the GOE distribution (solid gray line) rather than the Poisson distribution (dashed gray line) expected for integrable systems.

One quantitative measure of how well RMT statistics describe the eigenvalue spacings is the average ratio of consecutive spacings,
\begin{equation}
\tilde{r}_i \equiv \min\Big(\frac{s_i}{s_{i-1}},\frac{s_{i-1}}{s_i}\Big),
\end{equation}
which was studied in~\cite{Srdinsek:2020bpq} for the double Sine-Gordon model. One advantage of this quantity is that it does not require unfolding. The RMT prediction is $\langle \tilde r\rangle_{\rm GOE}\approx0.536$ compared to the Poisson prediction $\langle \tilde r\rangle_{\rm P}\approx0.386$. From our LCT data at $\lambda=1.6m^2$ we obtain $\langle \tilde r\rangle\approx0.535$ at $\Delta=40$ and $\langle \tilde r\rangle\approx0.530$ at $\Delta=45$ (again restricting to $1<\varepsilon/m^2<100$).

While this confirms that $\phi^4$ theory features RMT physics at strong coupling, an obvious question remains: at which value of the coupling does the spectrum transition from Poisson to GOE (i.e.~integrable to chaotic)? A convenient diagnostic for this crossover is the ratio of the distances of $P(s)$ from the two distributions,
\begin{equation}
\eta \equiv \frac{||P - P_{\textrm{Poisson}}||_1}{||P - P_{\textrm{GOE}}||_1} = \frac{\int_0^\infty ds|P(s) - e^{-s}|}{\int_0^\infty ds|P(s) - \frac{\pi}{2} s e^{-\frac{\pi}{4} s^2}|} \in [0,\infty],
\end{equation}
such that $\eta \ll 1$ corresponds to Poisson and $\eta \gg 1$ corresponds to GOE Wigner-Dyson.

\begin{figure}
\centerline{
\begin{overpic}[width=0.9\linewidth,tics=10]{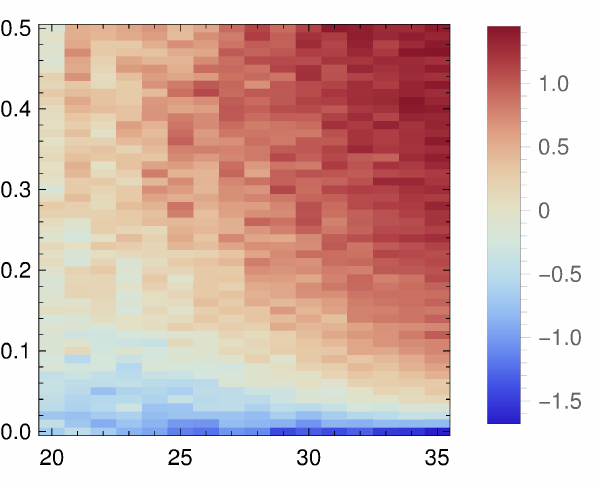}
	 \put (-5,41) {{\rotatebox{90}{$\lambda$}}} 
	 \put (37,0) {{$\Delta$}} 
	 \put (37,60) {{\bf \color{white}Wigner-Dyson}} 
	 \put (15,12) {\bf Poisson} 
	 \put (95,45) {{$\log\eta$}} 
\end{overpic}}
\caption{\label{fig_eta} Logarithm of the parameter $\eta$ defined in Eq.~\eqref{fig_eta}, for couplings $0\leq \lambda \leq 0.5$ and $\Delta=20,21,\ldots ,35$. The red regions ($\log\eta>0$) are closer to Wigner-Dyson, and the blue ($\log\eta<0$) to Poisson. }
\end{figure}
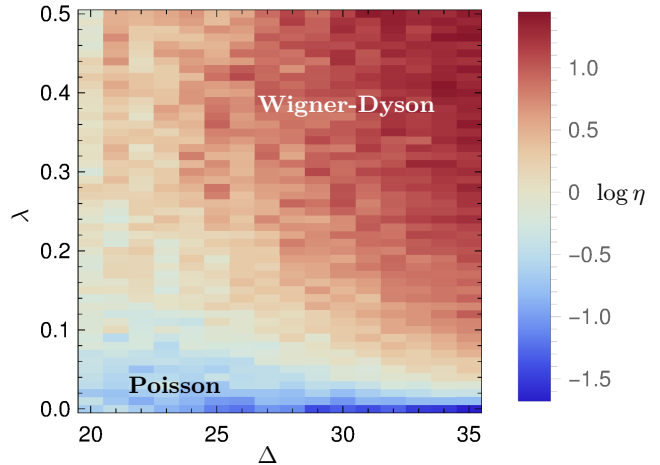

Fig.~\ref{fig_eta} shows $\log(\eta)$ for couplings $0\leq \lambda/m^2 \leq 0.5$ ($y$-axis) at various levels of truncation $20 \leq \Delta \leq 35$ ($x$-axis). At low $\Delta$, it is unclear which distribution describes these weakly-coupled theories. However, as we increase the size of our basis, all of these couplings (apart from $\lambda=0$) become better described by GOE ($\log \eta > 0$). This indicates that $\phi^4$ theory is chaotic for any nonzero coupling and deviations from GOE Wigner-Dyson at weak coupling are due to finite truncation effects.

\subsubsection*{Eigenvector Coefficients}

In addition to the eigenvalues, we also expect the components of eigenvectors in a given energy band to be well-described by RMT. Following~\cite{PhysRevE.81.036206,PhysRevE.85.036209}, we consider the overlap of energy eigenstates $|E\>$ with basis states $|i\>$,
\begin{equation}
c_{Ei} \equiv \<E|i\>.
\end{equation}
In our case, the basis states are the free scalar CFT states.
The GOE prediction for these coefficients is~\cite{RevModPhys.53.385,Alhassid:2000hf}
\begin{equation}
P_{\textrm{GOE}}(|c|^2) = \frac{1}{\sqrt{2\pi\sigma|c|^2}} e^{-\frac{|c|^2}{2\sigma}},
\label{eq:CoeffRMT}
\end{equation}
such that $\int_0^\infty dx \, P(x) = 1$ and $\int_0^\infty dx \, x P(x) = \sigma$.
This is a particularly appealing probe of chaos as it can test for thermality of an individual eigenstate---by studying the statistics of $\langle i | E\rangle$ with $|E\rangle$ fixed---in contrast to eigenvalue statistics which tests for thermality of a region of the spectrum.

In~\cite{Srdinsek:2020bpq}, it was found that the eigenstate coefficients do \emph{not} match the RMT prediction~\eqref{eq:CoeffRMT} in multiple 1+1d QFTs, including $\phi^4$ theory. Specifically, the tails of the distribution at small $|\<E|i\>|^2$ decay polynomially, rather than exponentially. While our LCT results reproduce this behavior, we find that this is a consequence of considering the overlaps of eigenstates with \emph{all} basis states. If we instead restrict to basis states whose energy expectation values $\<i|H|i\>$ are close to $E$, this polynomial behavior is eliminated, resulting in good agreement with RMT. Our interpretation of the need to restrict the average energy of the basis states is that the full Hamiltonian is not a truly random matrix and at large separation of energy scales the eigenstate coefficients are sensitive to theory-dependent dynamics. A more detailed discussion of this observation is presented in appendix~\ref{app_evecstatistics}.

\begin{figure}
\centerline{
\begin{overpic}[width=0.9\linewidth,tics=10]{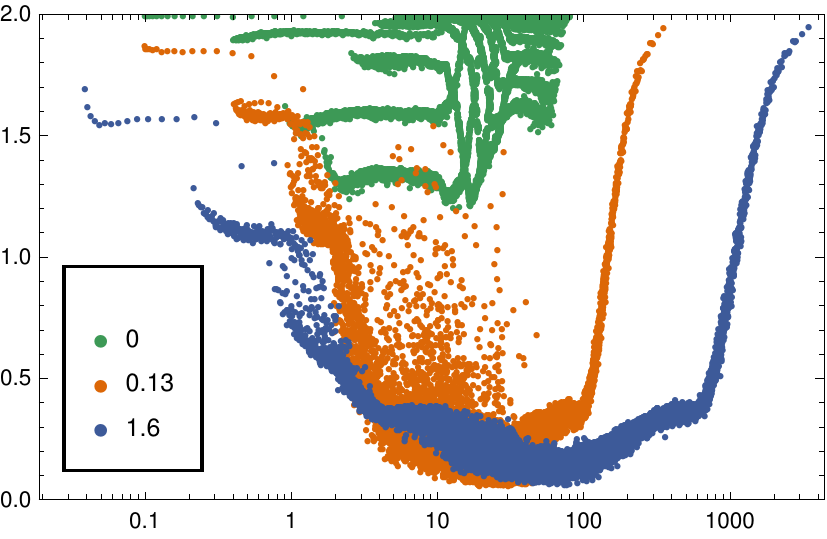}
	 \put (-5,25) {{\rotatebox{90}{$||P-P_{\rm GOE}||_1$}}} 
	 \put (49,-2) {{$\varepsilon/m^2$}} 
	 \put (11,28) {{$\lambda/m^2$}} 
\end{overpic}}
\caption{\label{fig_evecstats} Deviation of the coefficient statistics $P(|\langle E | i\rangle|^2)$ for individual eigenstates $|E\rangle$, from the GOE prediction. For each eigenstate $|E\rangle$, only one third of the coefficients $|\langle E|i \rangle|^2$ corresponding to the basis states with average energy $\langle i|H|i\rangle$ closest to $E$ were kept in the statistics. At strong coupling $\lambda=1.6m^2$, all eigenstates in the middle of the spectrum have close to random wavefunctions in the computational basis; at weak coupling $\lambda=0.13m^2$ there are outliers.}
\end{figure}

\begin{figure}
\centerline{
\begin{overpic}[width=0.9\linewidth,tics=10]{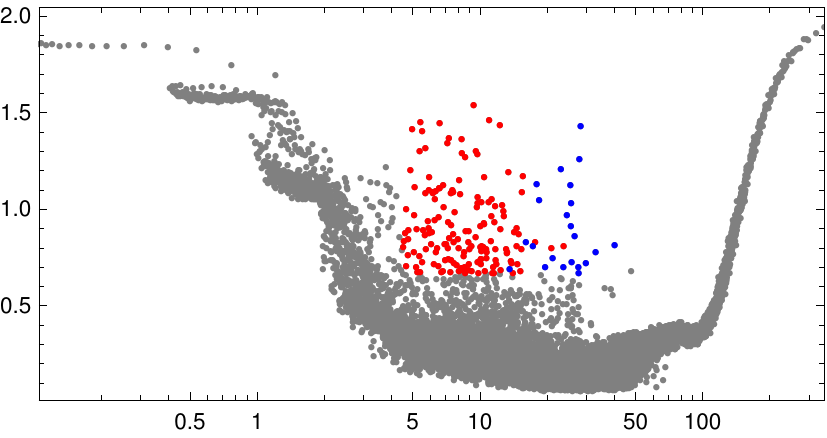}
	 \put (-5,15) {{\rotatebox{90}{$||P-P_{\rm GOE}||_1$}}} 
\end{overpic}}
\vspace{-11pt}
\centerline{
\begin{overpic}[width=0.9\linewidth,tics=10,trim={-0.06cm 0cm 0cm 0cm},clip]{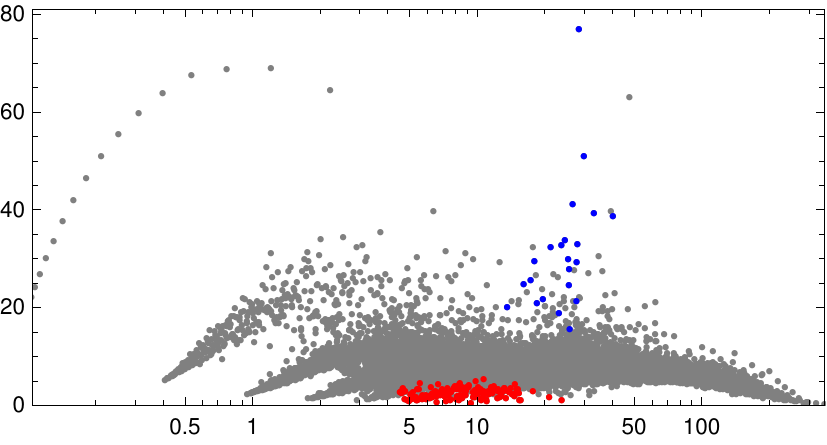}
	 \put (-5,15) {{\rotatebox{90}{$ \langle E | (\partial^2 \phi)^2|E\rangle$}}} 
	 \put (47,-3) {{$\varepsilon/m^2$}} 
\end{overpic}}
\caption{\label{fig_scars} At weak coupling $\lambda = 0.13m^2$, the states with wavefunctions far from RMT (top, chosen using the criteria $4.5<\varepsilon/m^2<45$ and $||P-P_{\rm GOE}||_1>\frac23$) are seen to violate ETH (bottom), with expectation values clearly below (red) or above (blue) the microcanonical expectation value.}
\end{figure}

With this in mind, we can compute for each eigenstate $|E\rangle$ the distribution of coefficients $P(|\<E|i\>|^2)$ where only a fraction of basis states $|i\rangle$ with average energy $\langle i|H|i\rangle$ closest to $E$ are kept, and compare the result to the GOE distribution \eqref{eq:CoeffRMT}, with $\sigma$ computed from $P(|\<E|i\>|^2)$. We have specifically chosen to restrict to a fixed fraction (one-third) of the total number of basis states at a given $\Delta$, so that the same number of coefficients is kept for each eigenstate; see appendix~\ref{app_evecstatistics} for more details.

Fig.~\ref{fig_evecstats} shows the resulting deviation of the eigenvector coefficients from the GOE prediction as a function of the energy density $\varepsilon$ for a free ($\lambda=0$), weakly-coupled ($\lambda=0.13m^2$), and strongly-coupled ($\lambda=1.6m^2$) theory, obtained at $\Delta=40$. As expected, the free theory coefficients (green) are far from GOE, and we can clearly see multiple distinct sectors associated with states of different particle number. However, at both weak (orange) and strong (blue) coupling, we see a significant band of states which are well-described by RMT ($||P - P_{\textrm{GOE}}|| \rightarrow 0$). At large values of $\varepsilon$, the eigenvector coefficients begin to deviate significantly from GOE due to truncation effects, as their eigenvalues are near the effective UV cutoff set by $\Delta$. At small values of $\varepsilon$, the coefficients also deviate from GOE, due to the fact that these are largely non-thermal states with low particle number.

One notable feature of these results is that while the majority of states at weak coupling with $1 \lesssim \varepsilon/m^2 \lesssim 100$ are well-described by GOE, we can also clearly see so-called ``scar'' states which are not (see \cite{Serbyn:2020wys,Moudgalya:2021xlu,Chandran:2022jtd} for reviews on many-body quantum scars). Importantly, though, these scar states have large enough energy and particle number density that they are \emph{not} weakly-interacting. Fig.~\ref{fig_scars} shows that these same states violate ETH. These states are also visible in the coefficient entropy, discussed in appendix \ref{app_evecstatistics} and shown in Fig.~\ref{fig_evecstats_coeffentropy}. It would be interesting to study these particular states in more detail to determine whether they correspond to multi-particle states near threshold, which are expected to have a semiclassical description~\cite{Son:1995wz,Rubakov:1995hq,Badel:2019oxl}. We note that this potential mechanism for quantum many-body scars appears to be much more robust than other known mechanisms in the literature, which typically require delicate cancellations between decay channels 
	\footnote{Scars of a different nature were found in the symmetry broken phase of a 1+1d QFT using Hamiltonian truncation in Ref.~\cite{Robinson:2018wbx}. However the temperature considered there was no larger than the mass gap, so that one expects slow thermalization due to Boltzmann suppression of the equation of state. We are focusing on the strongly thermalizing regime in this work.}.

At strong coupling, however, we see that there are no scar states with $1 \lesssim \varepsilon/m^2 \lesssim 300$, and all coefficients in this range follow a narrow band well-described by GOE. This shows that at strong coupling all states away from the edges of the spectrum look thermal, as expected in generic interacting quantum many-body systems in the absence of disorder \footnote{One exception to this expectation are systems with dipole conservation \cite{Sala:2019zru}. However, the associated symmetries are usually not robust and can be broken by relevant deformations---they will therefore typically not emerge in systems that do not already have them microscopically.}.

\section{Dynamics}

\subsubsection*{Spectral Form Factor}

\begin{figure}
\centerline{
\begin{overpic}[width=0.9\linewidth,tics=10]{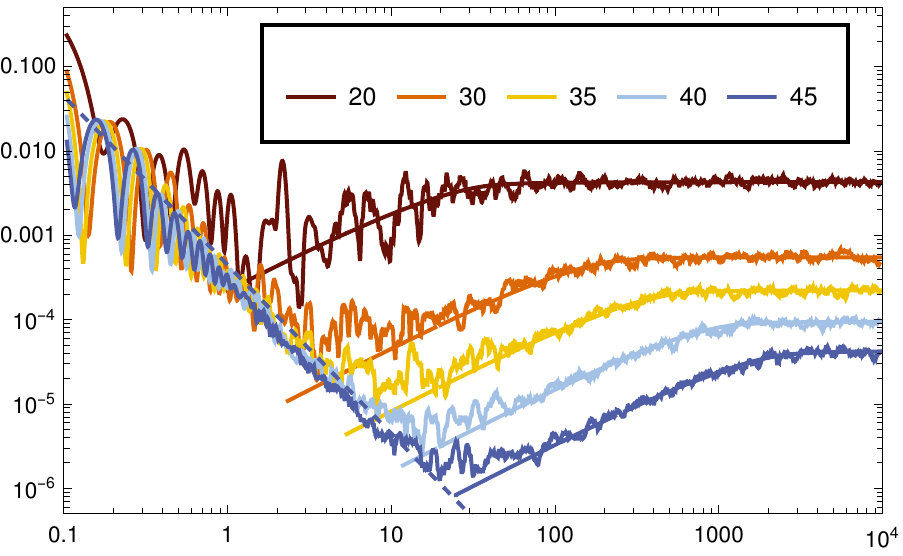}
		 \put (-5,27) {\scalebox{0.9}{\rotatebox{90}{${\rm SFF}(t)$}}} 
		 \put (58,54) {\scalebox{1}{$\Delta$}} 
		 \put (54,-2) {\scalebox{1.1}{$t$}} 
	\end{overpic}
} 
\caption{\label{fig_SFF} Time averaged spectral form factor \eqref{eq_SFF}, for $\mathbb Z_2$-even states ($\lambda = 1.6 m^2$). All states with energy density $1<\varepsilon = \tfrac{E^2}{\Delta}<100$ are kept in the microcanonical sum. The solid lines are the ramp and plateau expected from RMT for GOE \eqref{eq_RMT_ramp}. The initial slope is compared to the theoretical prediction $\sim \frac{1}{t^2} \frac{1}{\Delta}$ (dashed line, only $\Delta=45$ shown for clarity) obtained from a smooth density of states, Eq.~\eqref{eq_SFF_early}.}
\end{figure}

Hamiltonian truncation directly gives access to real time dynamics of quantum field theories, allowing the observation of thermalization of these systems at finite temperature. A standard probe of thermalization is the equilibrium two-point function of a local operator
\begin{equation}
\langle \mathcal{O}(t) \mathcal{O} \rangle_\beta
	= \frac{1}{N^2}\sum_{E,E'} e^{i(E-E') t} |\langle E | \mathcal{O} | E'\rangle|^2\, , 
\end{equation}
where we have represented the thermal expectation value microcanonically by summing over all eigenvalues $E,E'$ in a window containing $N$ states centered at energies corresponding to the inverse temperature $\beta$. To make connection with the eigenvalue statistics in Sec.~\ref{sec_stats}, we consider a slightly simpler quantity that does not involve the matrix elements of an operator: the spectral form factor (SFF) \cite{berry1985semiclassical,Cotler:2016fpe}:
\begin{equation}\label{eq_SFF}
{\rm SFF}(t)
	= \frac{1}{N^2}\sum_{E,E'} e^{i(E-E') t}\, .
\end{equation}
The result, with sum computed at fixed $\Delta$ and averaged over time, is shown in fig.~\ref{fig_SFF}. At early times, the SFF is dominated by its disconnected part 
\begin{equation}
{\rm SFF_{disc}}(t) = |\overline{Z(t)}|^2\, , 
	\quad Z(t) = \frac{1}{N}\sum_{E} e^{iE t}\, , 
\end{equation}
where the absolute value is taken after performing the time averaging. This disconnected piece is well approximated by taking the density of states to be smooth. The Cardy density of states \eqref{eq_Cardy} together with the volume of states \eqref{eq_volume} implies that the average density of state at fixed $\Delta$ is constant in energy $\rho(E,\Delta) \approx e^{\sqrt{E V}} \sim e^{\sqrt{\Delta}}$. One then finds 
\begin{equation}\label{eq_SFF_early}
Z(t) \simeq \frac{\int_{E_{\rm min}}^{E_{\rm max}} e^{iE t} dE}{E_{\rm max}-E_{\rm min}}
	= \frac{e^{iE_{\rm max} t} - e^{iE_{\rm min} t}}{i(E_{\rm max}-E_{\rm min})t}\, , 
\end{equation}
leading to the observed $1/t^2$ behavior of the slope of the SFF at early times in Fig.~\ref{fig_SFF}.

At late times, the SFF is instead dominated by its connected part, and is commonly used as a test of quantum chaos: quantum chaotic systems are expected to have a late time averaged SFF that agrees with that of a random matrix. The SFF for random matrices has a distinctive ramp-plateau feature (or `correlation hole'), which for the GOE universality class is given by \cite{mehta2004random}
\begin{equation}\label{eq_RMT_ramp}
{\rm SFF}(t)
	= \frac{1}{N}\cdot\begin{cases}
     2 \frac{t}{t_{\rm H}} - \frac{t}{t_{\rm H}} \log \left(1+2 \frac{t}{t_{\rm H}}\right)& \text{for } t\leq t_{\rm H} \\
    2-\frac{t}{t_{\rm H}}\log \left(\frac{1+\frac{t_{\rm H}}{2t}}{1-\frac{t_{\rm H}}{2t}}\right) & \text{for } t> t_{\rm H} \, . 
  \end{cases}
\end{equation}
where $N \sim e^{\sqrt{\Delta}}$ is the number of eigenstates and $2\pi/t_{\rm H} \sim e^{-\sqrt{\Delta}}$ is the mean eigenvalue spacing in the microcanonical window (the time scale $t_H$ thus defined is the Heisenberg time). Fig.~\ref{fig_SFF} shows excellent agreement of the SFF with this RMT prediction at late time.

\subsubsection*{Symmetry-resolved Spectral Form Factor}

\begin{figure}
\centerline{
\subfigure{
\begin{overpic}[width=0.95\linewidth,tics=10]{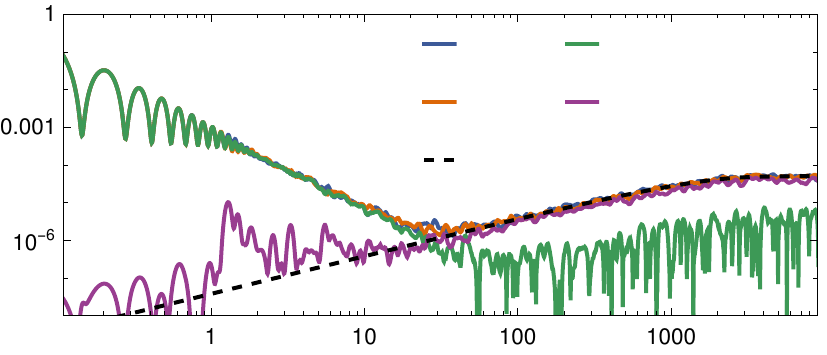}
	 \put (15,36) {\scalebox{0.8}{$\Delta=45,\, \lambda/m^2=1.6$}} 
	 \put (57,37) {\scalebox{0.8}{$S_{++}$}} 
	 \put (57,30) {\scalebox{0.8}{$S_{--}$}} 
	 \put (74,37) {\scalebox{0.8}{$S_{+-}$}} 
	 \put (74,30) {\scalebox{0.8}{$\frac{S_{--}+S_{++}}2 - S_{+-}$}} 
	 \put (57,22.5) {\scalebox{0.8}{GOE}} 
\end{overpic}
}}\vspace{-24pt}
\centerline{
\subfigure{
\begin{overpic}[width=0.95\linewidth,tics=10]{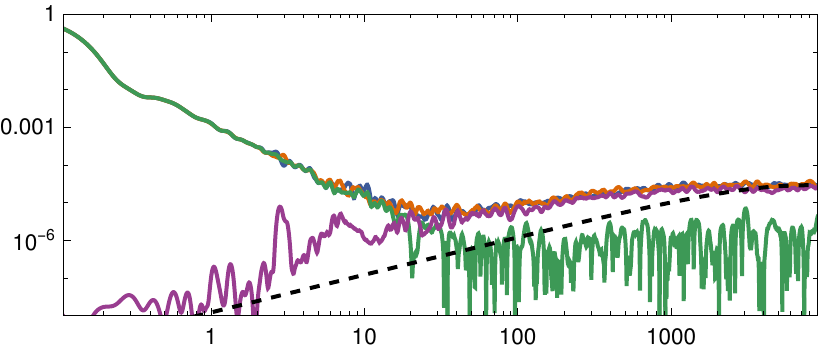}
	 \put (15,34) {\scalebox{0.8}{$\Delta=45,\, \lambda/m^2=0$}} 
\end{overpic}
}}\vspace{-24pt}
\centerline{
\subfigure{
\begin{overpic}[width=0.95\linewidth,tics=10]{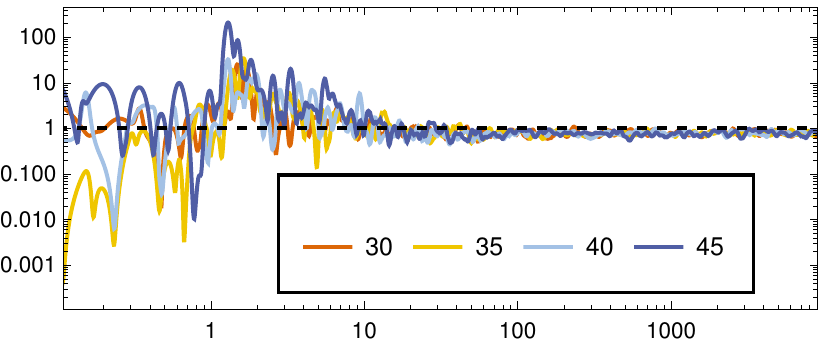}
	 \put (60,32) {\scalebox{0.9}{$\lambda/m^2=1.6$}} 
	 \put (60,16) {\scalebox{0.9}{$\Delta$}} 
	 \put (53,-3) {$t$} 
\end{overpic}
}}
\caption{\label{fig_app_sff} Symmetry resolved spectral form factors \eqref{eq_sff_sym} for states with energy density $10\leq \varepsilon/m^2 \leq 100$, time-averaged to remove noise. At strong coupling (top), the combination \eqref{eq_SFF_Z2_connected} (purple) agrees with the RMT prediction for the connected SFF down times as early as $t\sim 10 \frac{1}{m}$. In the free theory (center), the SFF is not that of a random matrix, but still exhibits a small correlation hole. The bottom figure shows the deviation of the combination \eqref{eq_SFF_Z2_connected} from the RMT prediction: $[\frac12({S_{--}+S_{++}}) - S_{+-}]/{\rm SFF_{GOE}}(t)$, for several values of $\Delta$ at strong coupling.
}
\end{figure}
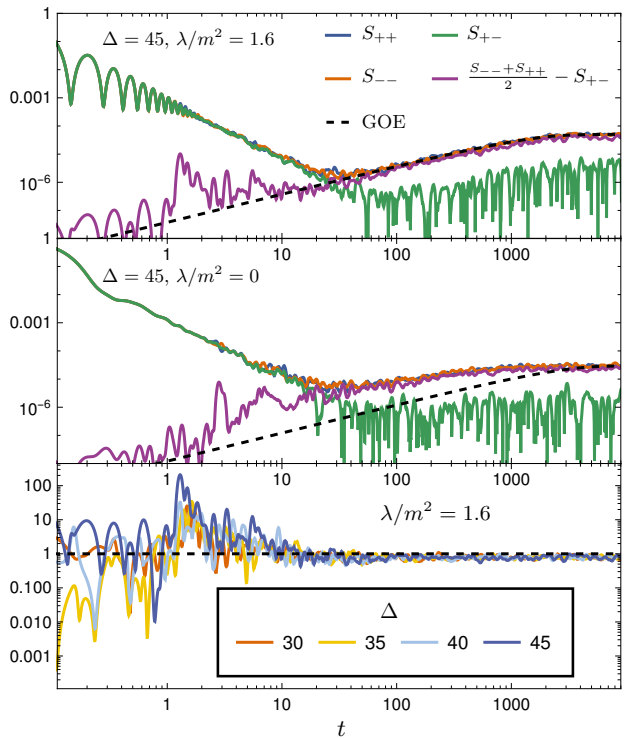

When global symmetries are present, RMT is only expected to describe the spectrum within superselection sectors. For example, the spectral form factor (SFF) shown in Fig.~\ref{fig_SFF} agrees with the RMT prediction because only $\mathbb Z_2$-even states are considered, and all states have the same momentum (due to exact Lorentz invariance all momentum sectors are identical in our approach). It is interesting to further make use of the global symmetry, here $\mathbb Z_2$, to construct symmetry-resolved spectral form factors. Consider
\begin{subequations}\label{eq_sff_sym}
\begin{align}
S_{++}(t) &= Z_+(t)Z_+^*(t)\, , \\
S_{--}(t) &= Z_-(t)Z_-^*(t)\, , \\
S_{+-}(t) &=\tfrac12 \left(Z_+(t)Z_-^*(t)+ Z_-(t)Z_+^*(t)\right)\, ,
\end{align}
\end{subequations}
where $Z_+(t) = \frac{1}{N_{+}} \sum_{E_+} e^{i E_+ t}$ is the partition function in the $\mathbb Z_2$-even sector, and $Z_-(t)$ that in the $\mathbb Z_2$-odd sector. These are shown in Fig.~\ref{fig_app_sff}. Both $S_{++}$ and $S_{--}$ are regular SFFs, in the $\mathbb Z_{2}$-even and odd sectors respectively, and exhibit the slope-ramp-plateau profile as expected. Instead, $S_{+-}$ only features the slope, fixed by the equation of state: the plateau is absent because no two eigenvalues of opposite $\mathbb Z_2$ parity are expected to match, and the ramp is absent because eigenvalues with different quantum numbers do not repel.

This observation allows for the construction of a linear combination of symmetry-resolved spectral form factors, where the slope approximately cancels:
\begin{equation}\label{eq_SFF_Z2_connected}
\frac12 \left(S_{++}(t) + S_{--}(t)\right) - S_{+-}(t)\, .
\end{equation}
The cancelation of the slope, which is controlled by the density of states, relies on the fact that the density of states is approximately $\mathbb Z_2$-symmetric. Fig.~\ref{fig_app_sff} shows that the linear combination \eqref{eq_SFF_Z2_connected} indeed removes the slope to very high precision. This procedure therefore allows the evaluation of a form of {\em connected} SFF, without the need of disorder averaging. This is particularly appealing in theories that do not have a natural large set of couplings that one can average over; however it may also be useful for disordered theories, as disorder averages can be costly. We expect symmetry-resolved SFFs \eqref{eq_sff_sym}, \eqref{eq_SFF_Z2_connected} to be useful probes of chaos in future simulations of quantum many-body systems.

It is tempting to interpret the mild excess of the connected SFF above the GOE prediction at intermediate times $1\lesssim t m \lesssim 10$ (Fig.~\ref{fig_app_sff} top) as possible evidence of suppression of spectral rigidity due to hydrodynamics, following \cite{Friedman:2019gyi,Winer:2020gdp} (see discussion in Sec.~\ref{sec_discuss}). However, another possible origin is inexact cancellation between the $\mathbb Z_2$-even and odd density of states in \eqref{eq_SFF_Z2_connected}. With the current resolution we have not been able to unambiguously establish the origin of this feature. The bottom panel in Fig.~\ref{fig_app_sff} shows the dependence of this feature on system size, through $V = \Delta /E= \sqrt{\Delta/\varepsilon}$. 

\subsubsection*{Survival Probability}

As a final measure of real time dynamics, we will consider the `survival probability'
\begin{equation}\label{eq_survival}
W_i(t) \equiv |\langle i(0)|i(t) \rangle|^2 = \sum_{EE'} |c_{iE}|^2|c_{iE'}|^2 e^{i(E-E')t}\, ,
\end{equation}
which probes the thermalization of a basis state $|i\rangle$ (see, e.g., \cite{Torres-Herrera:2017qwo}). While in this work we compute the survival probability by exact diagonalization of the truncated Hamiltonian, an advantage of this observable is that it can also be obtained by time evolution of the basis state through other means (e.g.~\cite{PhysRevB.99.094419}). The analogy with the SFF \eqref{eq_SFF} is clear---the survival probability can be viewed as `dynamical typicality' applied to the SFF \cite{PhysRevLett.112.120601}. Conversely, the SFF is the survival probability for the thermofield double state \cite{delCampo:2017bzr}.

\begin{figure}
\centerline{
\begin{overpic}[width=0.9\linewidth,tics=10]{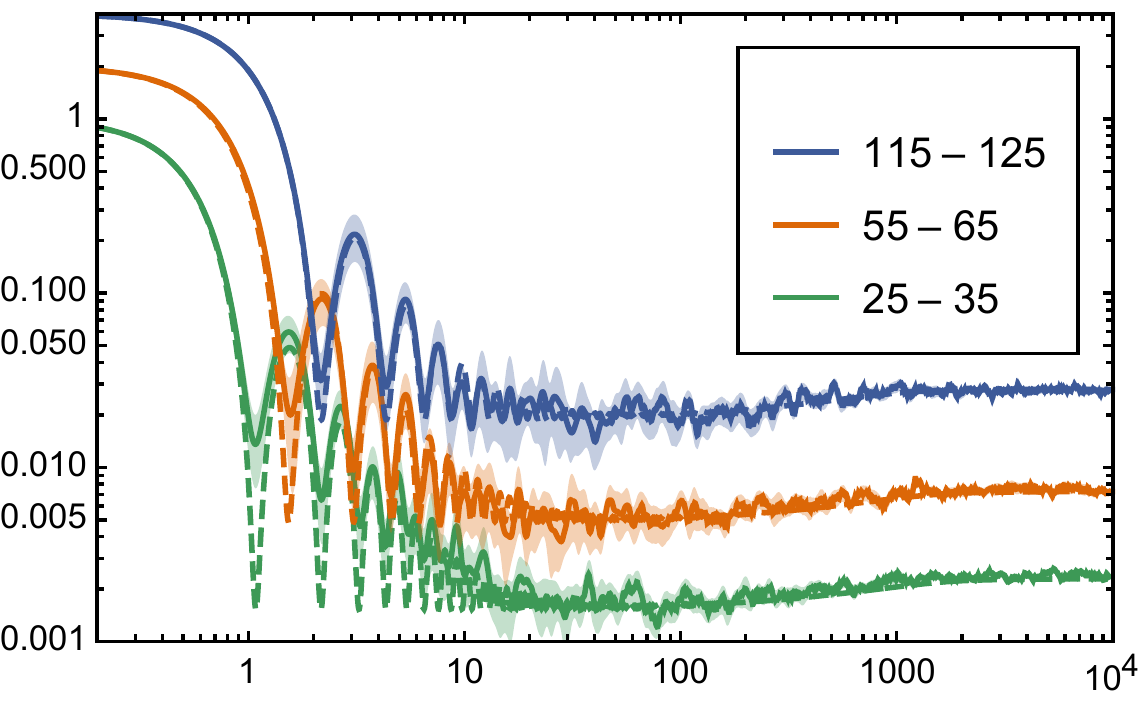}
		 \put (-5,27) {\scalebox{0.9}{\rotatebox{90}{${W}(t)$}}} 
		 \put (77,52) {\scalebox{0.9}{$\varepsilon / m^2$}} 
		 \put (54,-2) {\scalebox{1.1}{$t$}} 
	\end{overpic}
} 
\caption{\label{fig_survival} Survival probability for basis states with energy density in three different windows (y-axes are offset for clarity), obtained at $\Delta=40$, $\lambda = 1.6m^2$. The solid line denotes the mean result for the twenty random states selected from each window, and shaded region the standard deviation. The dashed lines are the predictions from the density of states (slope at early times) and RMT (ramp and plateau at late times).}
\end{figure}

To reduce noise, we average \eqref{eq_survival} over several basis states with average energies $ \langle i|H|i\rangle$ that lie in a somewhat narrow window. The result at strong coupling is shown in Fig.~\ref{fig_survival} for three different energy windows: $\varepsilon/m^2 \in 25$--$35$, $55$--$65$, $115$--$125$. The basis states in these three windows have an average particle number of $7$, $9$, $12$, and an average scaling dimension of $35$, $36$, $35$, respectively. The LCT results in Fig.~\ref{fig_survival} are compared to the RMT prediction (dashed lines)~\cite{mehta2004random,Torres-Herrera:2017qwo}, which is structurally the same as for the SFF. At early times, it is dominated by the disconnected piece proportional to Eq.~\eqref{eq_SFF_early} coming from the density of states, resulting in the same $1/t^2$ slope. At late times, it is dominated by the connected piece proportional to Eq.~\eqref{eq_RMT_ramp}, leading to a similar ramp-plateau structure, though with a shallower correlation hole.

\section{Discussion}\label{sec_discuss}

The results above establish that a paradigmatic non-integrable 1+1d QFT thermalizes and displays chaos, as manifested by spectral statistics and late time dynamics following RMT. While this is the expectation for generic quantum many-body systems, most analytically tractable systems are not generic, and most numeric studies have been performed on lattices; this work suggests that continuum theories, whose microscopic structure---UV CFTs---are entirely different than a lattice, also follow this expectation. Furthermore, theoretical control over certain limits of this theory, and in particular its asymptotic freedom, allowed for quantitative tests of thermalization: the expectation value of simple operators in macroscopic high energy eigenstates agrees with thermal expectation values in the UV CFT, when the energy density of these states is large compared to the dimensionful scales of the QFT.

The Hamiltonian of a QFT of course has a rich structure far beyond that of a random matrix. In particular, locality of the underlying system should lead to the emergence of hydrodynamics at intermediate times; long-wavelength modulations of hydrodynamic densities act as approximately conserved quantities that suppress spectral rigidity between eigenvalues of the Hamiltonian at energy differences larger than the Thouless energy \cite{Friedman:2019gyi}. In diffusive systems of linear size $L$ with diffusivity $D$, the Thouless energy is $E_{\rm Th} = D/L^2$---however, hydrodynamics in 1+1d QFTs is expected to be described by the KPZ universality class rather than diffusion, where instead $E_{\rm Th} = \mathcal D / L^{3/2}$ with $\mathcal D$ entirely fixed by the equilibrium equation of state \cite{Delacretaz:2021ufg}. This loss of spectral rigidity can be probed through the spectral form factor \cite{Gharibyan:2018jrp}: at times $t$ between the local equilibration time $\tau_{\rm eq}$ and the Thouless time $\tau_{\rm Th} \equiv 1/E_{\rm Th}$, densities with wavevectors $k\lesssim 1/(\mathcal D t)^{1/z}$ have not fully decayed  and can be used as approximate quantum numbers ($z=3/2$ for KPZ and $z=2$ for diffusion). In $d$ spatial dimensions, this leads to $N_{\rm sectors}(t)\sim \exp \left( \frac{L^d}{(\mathcal D t)^{d/z}}\right)$ approximately decoupled sectors, and a corresponding behavior in the connected spectral form factor ${\rm SFF}(t) \simeq N_{\rm sectors}(t) \frac{t}{e^{2S}}$ for $\tau_{\rm eq} \lesssim t \lesssim \tau_{\rm Th} = L^z/\mathcal D$ \cite{Friedman:2019gyi,Winer:2020gdp}. In Fig.~\ref{fig_SFF}, the early time behavior of the SFF is dominated by the disconnected part, with fall-off $\sim 1/t^2$ fixed by the equation of state. Removing the disconnected part is difficult without disorder averaging---the QFT under consideration \eqref{eq_S} has a single dimensionless coupling $\lambda/m^2$. The symmetry-resolved SFF introduced in Eq.~\eqref{eq_sff_sym} appears to efficiently capture the connected part of the SFF without averaging, and may be useful in detecting signatures of hydrodynamics, and future studies of quantum chaos in many-body systems more generally.

Emergence of hydrodynamics can also be searched for with a number of more refined observables, including equilibrium two-point functions, thermalization after a quench, entanglement growth, etc. The lower bounds on the local equilibration time $\tau_{\rm eq}$ of 1+1d QFTs \cite{Delacretaz:2021ufg} and the fact that these systems are expected to be superdiffusive further reduce the parametric window for hydrodynamic behavior; we nevertheless expect this regime to be accessible in the near future, and leave this important outstanding question for future work. Hydrodynamics in the KPZ universality class has yet to be observed numerically in non-integrable quantum many-body systems. 

On the lattice, a number of methods have been developed in the past decade to study quantum dynamics without exactly diagonalizing the Hamiltonian, thereby giving access to larger systems, see Ref.~\cite{RevModPhys.93.025003} for a recent review. We expect that our LCT approach can be similarly improved by developing or adapting techniques specifically geared towards finite temperature dynamics. A simple yet powerful example is dynamical typicality \cite{PhysRevLett.112.120601}, which consists in replacing the thermal state with a high (average) energy basis state. Assuming this state thermalizes, this method allows for the study of dynamics without diagonalization by solving Schr\"odinger equations numerically, and can be straightforwardly applied to LCT to compute e.g.~the survival probability \eqref{eq_survival}.

\begin{acknowledgments}
We wish to acknowledge helpful discussions with Cheryne Jonay, Jorrit Kruthoff, Anatoli Polkovnikov, Tibor Rakovszky and Brian Swingle. LVD is supported by the Swiss National Science Foundation
and the Robert R. McCormick Postdoctoral Fellowship of the Enrico Fermi Institute. LVD also acknowledges the hospitality of the the Aspen Center for Physics, supported by National Science Foundation grant PHY-1607611, where part of this work was completed. ALF and EK were supported in part by the US Department of Energy Office of Science under Award
Number DE-SC0015845, and in part by the Simons Collaboration Grant on the Non-Perturbative Bootstrap. MTW is partly supported by the National Centre of Competence in Research SwissMAP and the Swiss National Science Foundation. This work was completed in part with resources provided by the
University of Chicago Research Computing Center.
\end{acknowledgments}

\appendix

\section{Methodology}
\label{app:LCTReview}

The results in this work were obtained using the LCT Mathematica packages~\footnote{\href{https://github.com/andrewliamfitz/LCT}{https://github.com/andrewliamfitz/LCT}}, which implement the calculational method described in detail in~\cite{Anand:2020gnn}. Lightcone conformal truncation is a particular Hamiltonian truncation method which uses the basis of states from a UV CFT (in this work, the free massless scalar) which can be deformed to obtain the desired QFT (1+1d $\phi^4$ theory). The basis states are built from primary operators $\mathcal{O}$ in momentum space,
\begin{equation}
|\mathcal{O},P_-\> \equiv \frac{1}{N_\mathcal{O}} \int dx^- e^{-iP_-x^-} \mathcal{O}(x^-)|0\>,
\end{equation}
where $x^\pm \equiv \frac{1}{\sqrt{2}}(t \pm x)$ and $N_\mathcal{O}$ is an overall normalization factor. These states are formally defined in infinite volume, though, as discussed in appendix~\ref{app:SizeOfStates}, the resulting QFT energy eigenstates have finite volume resolution. For the free scalar, the necessary primary operators take the general form
\begin{equation}
\mathcal{O}= \sum_{\{k\}} C^\mathcal{O}_{\{k\}} \partial_-^{k_1} \phi \cdots \partial_-^{k_n} \phi.
\end{equation}

This basis is truncated to a finite-dimensional subspace by setting a maximum scaling dimension $\Delta$ and only keeping operators with $\Delta_\mathcal{O} \leq \Delta$. This is equivalent to restricting to operators with at most $\Delta$ derivatives. The QFT lightcone Hamiltonian
\begin{equation}
P_+ = \int dx^- \bigg( \frac{1}{2} m^2 \phi^2 + \frac{1}{4!} 4\pi\lambda \phi^4 \bigg),
\end{equation}
is evaluated in this subspace, and the resulting finite-dimensional matrix is numerically diagonalized to obtain the approximate QFT energy eigenstates $|\mu,P_-\>$.

In this work, the largest basis we consider is $\Delta = 45$, with $44\,601$ $\mathbb{Z}_2$-odd states and $44\,533$ $\mathbb{Z}_2$-even states. The most computationally prohibitive steps in this calculation are the construction of the Hamiltonian matrix elements for the $\phi^4$ interaction and the numerical diagonalization, which take approximately 2 hours and 6 hours on a typical laptop, respectively.

\section{Size of States}
\label{app:SizeOfStates}

\begin{figure*}
\centerline{
	\subfigure{\label{sfig_label1}
	\begin{overpic}[height=0.195\linewidth,tics=10]{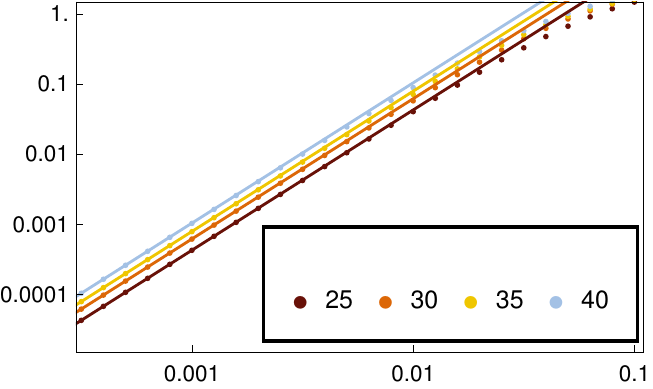}
		 \put (20,50) {(a)}
		 \put (-7,20) {\scalebox{0.9}{\rotatebox{90}{$f(q_-/P_-)$}}} 
		 \put (63,19) {\scalebox{1}{$\Delta$}} 
		 \put (54,-4) {\scalebox{1.1}{$q_-/P_-$}} 
	\end{overpic}
	}
\subfigure{\label{sfig_label2}
	\begin{overpic}[height=0.2\linewidth,tics=10]{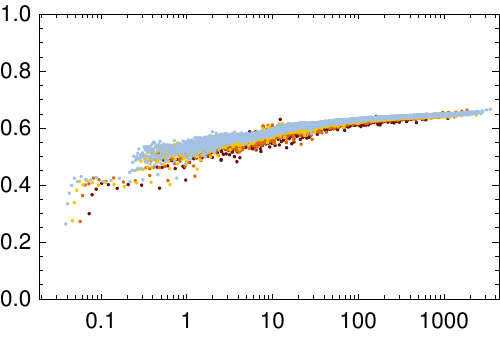}
		 \put (17,54) {(b)}
		 \put (-4,34) {\scalebox{1.1}{\rotatebox{90}{$\alpha$}}} 
		 \put (70,18) {\scalebox{1}{$\lambda=1.6$}} 
		 \put (54,-4) {\scalebox{1.1}{$\varepsilon$}} 
	\end{overpic}}\hspace{5pt}
\subfigure{\label{sfig_label3}
	\begin{overpic}[height=0.2\linewidth,tics=10]{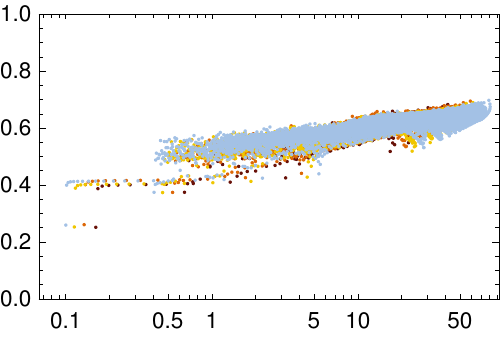}
		 \put (17,54) {(c)}
		 \put (-4,34) {\scalebox{1.1}{\rotatebox{90}{$\alpha$}}} 
		 \put (70,18) {\scalebox{1}{$\lambda=0$}} 
		 \put (54,-4) {\scalebox{1.1}{$\varepsilon$}} 
	\end{overpic}}
}
\caption{\label{fig_volume} (a) Form factor $f(q_-/P_-)\equiv 2 - \frac{1}{P_-} \langle\mu,P +q | T_{--} | \mu,P\rangle$ for a state with energy density $\varepsilon\simeq 10m^2$ at $\lambda=1.6$. The straight lines are fits to $\tilde \alpha^2 q_-^2/P_-^2$; this determines $\tilde \alpha$, which can be seen to increase with $\Delta$. (b) The coefficient $\alpha \equiv \tilde \alpha/\Delta$ is essentially independent of $\Delta$ and the energy of the state $\mu$, or its energy density $\varepsilon\equiv \mu^2/\Delta$. (c) The same holds in the free theory, albeit with a larger variance in the volume of states.
}
\end{figure*}
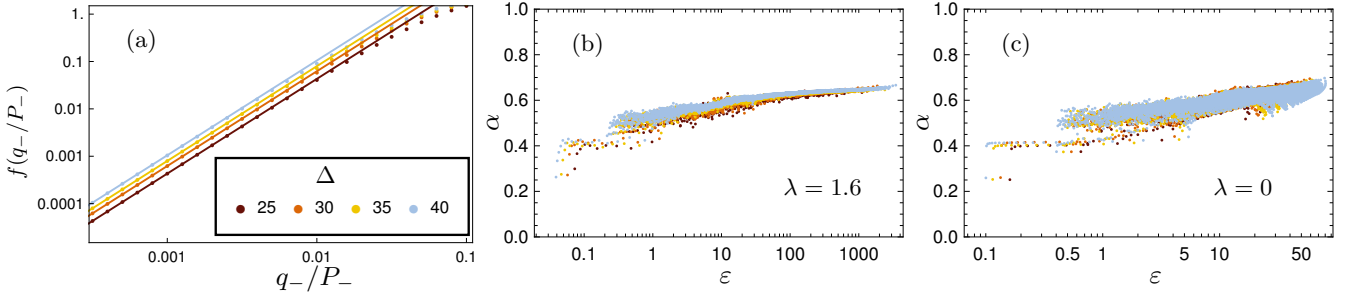

In exact diagonalization studies of lattice models, the system size is a fixed external parameter given by the number of sites. In certain Hamiltonian truncation approaches to QFT, the total volume is also a fixed parameter. In LCT, the volume is instead formally infinite, but the truncation introduces an IR (as well as UV) cutoff: in particular, the highly excited states that are being used as a thermal bath in this work have a finite size, which can be measured with a simple probe operator. In this appendix, we show how this measurement produces the result quoted in Eq.~\eqref{eq_volume} in the main text.

In a relativistic QFT in infinite volume, one can label states by their Lorentz-invariant momentum squared $\mu^2$, corresponding to the state's energy in its rest frame, and momentum $\vec P$. The spectrum at any $\vec P$ is simply related to that at any other  $\vec P'$ by boost symmetry.
To measure the volume (and shape) of a state $|\mu, \vec P \rangle$, one can probe it with a local operator---we will use the stress tensor, and consider its matrix elements
\begin{equation}
\langle \mu, \vec P + \delta \vec P | T_{\nu\lambda} | \mu, \vec P\rangle
\end{equation}
with a momentum transfer $\delta \vec P$. The dependence of this matrix element, or form factor, on $\delta \vec P $ indicates the shape and size of the state $|\mu\rangle$. Since all components of the stress tensor are related by Ward identities, we will focus on a specific one below: $T_{--}$
.

In light-cone quantization, conserved charges are obtained by integrating over $x^-$. One therefore has
\begin{equation}\label{eq_Pminus}
\int dx^- T_{--}(x^-,x^+) | \mu, P_-\rangle
	= | \mu, P_-\rangle P_-\, , 
\end{equation}
where $P_-$ denotes the momentum of the state in the $x^-$ direction. This equation is independent of $x^+$, by conservation. Using a Lorentz-invariant normalization of states
\begin{equation}
\langle \mu', P_-' | \mu, P_-\rangle
	= \delta_{\mu\mu'} 2\pi \delta (P_- - P_-') 2P_-\, , 
\end{equation}
one finds that \eqref{eq_Pminus} fixes the diagonal matrix elements of the stress tensor
\begin{equation}
\langle \mu, P_- | T_{--}(0)| \mu, P_-\rangle
	= 2P_-^2\, .
\end{equation}
For finite momentum transfers $q_- \equiv P'_- - P_- \neq 0$, the matrix element is no longer fixed. In the limit where this momentum transfer is small $q_-\ll P_-$, one can expand it as
\begin{equation}
\langle \mu, P_-+q_- | T_{--}(0)| \mu, P_-\rangle
	= 2P_-^2 \left(1 - \tilde \alpha^2 \frac{q_-^2}{P_-^2} + \cdots\right)
\end{equation}
A term $O(q_-)$ is absent: it is fixed similarly to the $O(1)$ term by using the fact that $\int dx^- T_{--}(x) x^-$ is the generator of boosts. The $O(q_-^2)$ term is the first that is not fixed by symmetry. Its coefficient is a measure of the square of the state's volume, in the $x^-$ direction: $V_-\sim \tilde \alpha/P_-$. Boosting to the states reference frame, one finds a volume
\begin{equation}
V \sim \frac{\tilde \alpha}{\mu}\, .
\end{equation}
This coefficient is measured in Fig.~\ref{fig_volume}. One finds that it scales with our truncation parameter as $\tilde \alpha = \Delta \alpha$, where $\alpha\sim 1$ is essentially independent of $\Delta$, and has a very weak average dependence on the eigenstate energy $\mu$.

\begin{figure}
\centerline{
\subfigure[$\lambda=0.13m^2$]{
\begin{overpic}[width=0.43\linewidth,tics=10]{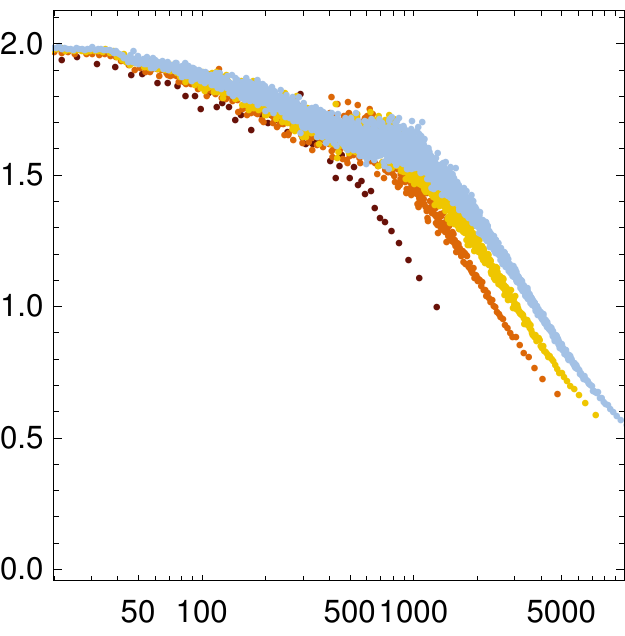}
	 \put (-10,28) {\scalebox{0.9}{\rotatebox{90}{$\langle E|\phi^2|E\rangle/E^2$}}} 
	 \put (50,-8) {\scalebox{0.9}{$E^2$}} 
\end{overpic}
\begin{overpic}[width=0.43\linewidth,tics=10]{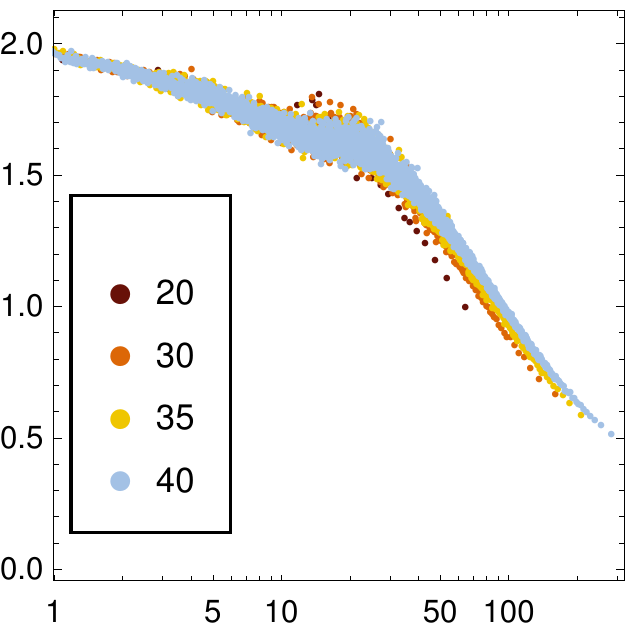}
	 \put (16,62) {\scalebox{0.8}{$\Delta$}} 
	 \put (50,-8) {\scalebox{0.9}{$E^2/\Delta$}} 
\end{overpic}
}}
\centerline{
\subfigure[$\lambda=1.6m^2$]{
\begin{overpic}[width=0.45\linewidth,tics=10]{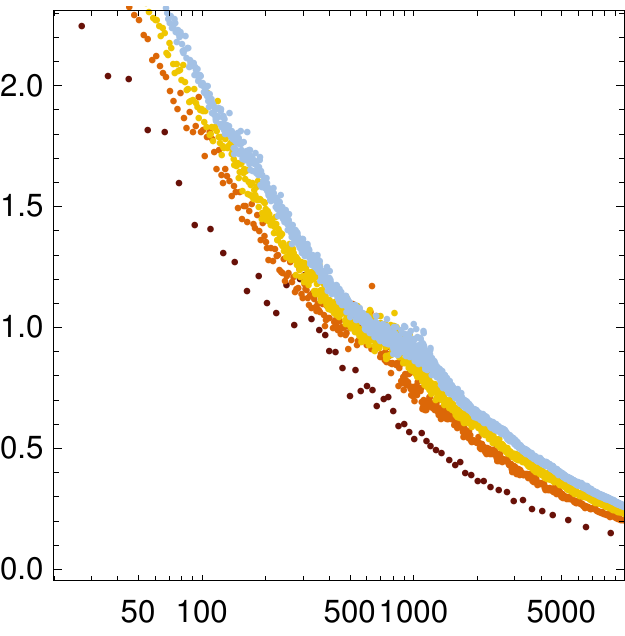}
	 \put (-10,28) {\scalebox{0.9}{\rotatebox{90}{$\langle E|\phi^2|E\rangle/E^2$}}} 
	 \put (50,-8) {\scalebox{0.9}{$E^2$}} 
\end{overpic}
\begin{overpic}[width=0.45\linewidth,tics=10]{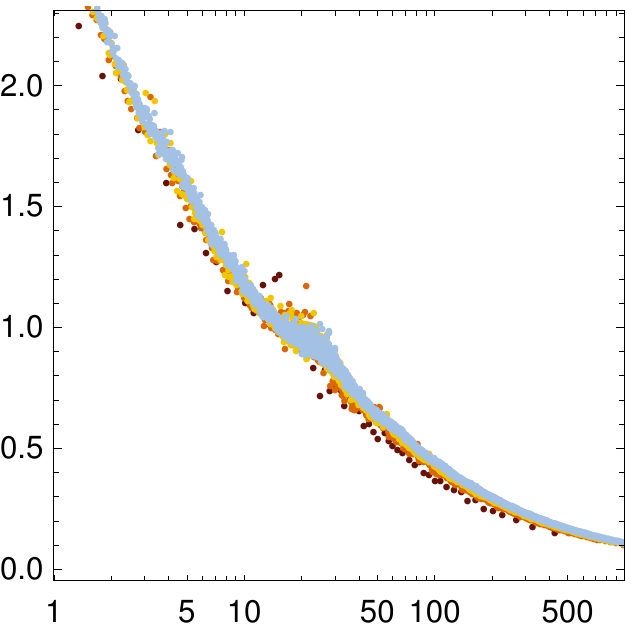}
	 \put (50,-8) {\scalebox{0.9}{$E^2/\Delta$}}
\end{overpic}
}}
\caption{\label{fig_collapse_app} 
Scaling collapse of $\phi^2$ expectation value at strong and weak coupling, when plotted against $E^2/\Delta$, as predicted by \eqref{eq_collapse}. 
}
\end{figure}
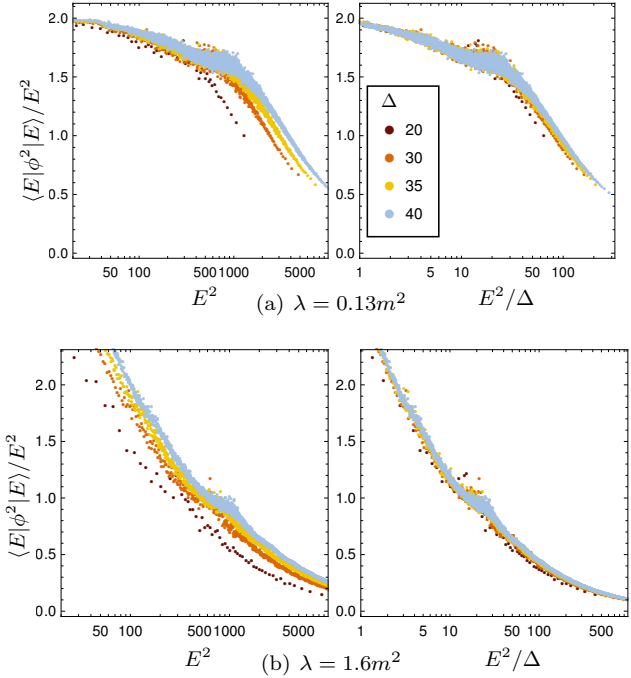

We therefore conclude that states with rest frame energy $\mu$ at a fixed truncation $\Delta$ have a volume
\begin{equation}\label{eq_app_volume}
V \sim \frac{\Delta}{\mu}\, .
\end{equation}
Therefore, the spectrum obtained numerically with LCT at a fixed $\Delta$ is not at a fixed volume: instead, the volume evolves smoothly following \eqref{eq_app_volume}. This has implications for canonical averages over states, and must be carefully acounted for in order to compare to more conventional (fixed volume) canonical ensembles, as discussed in Sec.~\ref{sec_thermo}.

From now on, we use the notation $E\equiv\mu$ to denote the rest mass energy of a state as in the main text. Eq.~\eqref{eq_app_volume} motivates defining the intensive energy density of an eigenstate $|E\rangle$ as
\begin{equation}\label{eq_app_epsilon}
\varepsilon = \frac{E}{V} = \frac{E^2}{\alpha \Delta}\, . 
\end{equation}
A non-trivial check of \eqref{eq_app_volume} is the scaling collapse observed in matrix elements of other local operators, when plotted against the energy density \eqref{eq_app_epsilon}. Figs.~\ref{fig_collapse} and \ref{fig_collapse_app} show such a scaling collapse for the operators $(\partial \phi)^4$ and  $\phi^2$, respectively.  Moreover, we expect that these matrix elements should be related to thermal expectation values after compensating for the finite volume of the states, as follows. First, we boost from the frame $p_-=1$  in which we usually work to the rest frame $p_- = \frac{E}{\sqrt{2}}$, so $\<E| \CO | E \>_{\rm rest} = (E/\sqrt{2})^{-J} \< E | \CO | E\>$, where $J$ is the spin of the operator $\CO$ ($J\equiv -1$ for $\partial_- \cong p_-$).  Then we compensate for the fact that the states are plane-wave normalized, which in the rest frame can be written as $\< E | E\> \equiv (2 \pi) 2 E \delta(p_x - p_x)$. Because of the plane-wave normalization, the normalized expectation values $\frac{\< E | \CO | E\>}{\< E | E \>}$ 
vanish for local observables $\CO$ -- in physical terms, our states $E$ have a finite size whereas the average is being taken over all of space, which is mostly vacuum.  Physical operator densities can be obtained with the replacement $2 \pi \delta(p_x -p_x) \rightarrow V_x = \frac{\alpha \Delta}{E}$:
%
\begin{equation}
\< \CO \>_E \equiv \frac{(2\pi)  \delta(p_x - p_x)}{V_x}  \frac{E^{-J} \< E | \CO | E\>}{\< E| E\>}  = \frac{\< E | \CO | E\>}{\alpha E^J \Delta}.
\end{equation}
We expect this quantity to be proportional to a thermal expectation value of the density of $\CO$. 
%
%
Comparing with our definitions in (\ref{eq_collapse}), and using $\varepsilon = \frac{E^2}{\alpha \Delta}$, we see that
\begin{equation}
\tilde{f}_\CO(\varepsilon) = \alpha^{\frac{J}{2}+1} \varepsilon^{J/2} \< \CO \>_E.
\end{equation}

\section{Theory Considerations}\label{app_theory}

\subsubsection*{High Temperature Behavior}

The high temperature regime of general 1+1d QFTs was studied in~\cite{Delacretaz:2021ufg} using conformal perturbation theory to derive strong lower bounds on the equilibration time $\tau_{\textrm{eq}}$ characterizing the emergence of hydrodynamics. However, it was shown that this naive conformal perturbation theory approach breaks down for $\phi^4$ theory, due to the lack of a thermal mass for a 1+1d massless scalar. As a result, the high temperature equation of state is non-analytic in the coupling.

For example, in free massive scalar field theory at $T \gg m$ the pressure is
\begin{equation}
P = \frac{\pi}{6} T^2 - \frac{1}{2} mT + \ldots
\end{equation}
such that the dimensionless entropy density $s_o \equiv \frac{s}{T}$ is
\begin{equation}
s_o = \frac{\pi}{3} - \frac{m}{2T} + \ldots
\end{equation}
With the addition of a $\phi^4$ interaction, at $T \gg \sqrt{\lambda}$ we instead have
\begin{equation}
\begin{aligned}
P &= \frac{\pi}{6} T^2 - a_0 \bigg(\frac{4\pi}{4!} \lambda T \bigg)^{1/3} T + \ldots, \\
s_o &= \frac{\pi}{3} - \frac{4}{3} a_0 \bigg(\frac{4\pi}{4!} \frac{\lambda}{T^2} \bigg)^{1/3} + \ldots,
\end{aligned}
\end{equation}
with $a_0 \approx 0.667986$~\cite{Delacretaz:2021ufg,Janke:1995zz}. In both cases, we see that $s_o$ is monotonically decreasing as we decrease $T$.

At late times, the high temperature behavior of 1+1d $\phi^4$ theory is described by the KPZ universality class, whose dissipation is entirely fixed by thermodynamics, leading to the following causality bound on the equilibration time (see~\cite{Delacretaz:2021ufg} for more details)
\begin{equation}
\tau_{\textrm{eq}} \gtrsim \frac{1}{T} \bigg( \frac{T^2}{\lambda} \bigg)^{1/3} \gg \frac{1}{T}.
\end{equation}

\subsubsection*{Free Thermal Expectation Values}

The leading high temperature expectation values for local operators are those of the free scalar CFT, which are related to vacuum correlation functions via a conformal transformation~\cite{Cardy:1986ie}. As a result, the only operators with nonzero expectation values are those built from the stress tensor (i.e.~Virasoro descendants of the vacuum).

For example, the stress tensor component $T_{--} \equiv (\partial_-\phi)^2$ has the thermal expectation value
\begin{equation}
\<(\partial_-\phi)^2\> = \frac{\pi}{6} T^2 \equiv \varepsilon.
\end{equation}
The next operator with a nonzero thermal expectation value is the Virasoro descendant
\begin{equation}
T_{--}^2 + \frac{3}{20\pi} \partial_-^2 T_{--} = (\partial_-\phi)^4 + \frac{1}{2\pi} \Big( (\partial_-^2 \phi)^2 - \frac{1}{5} \partial_-^2(\partial_-\phi)^2 \Big).
\end{equation}
There is also a Virasoro primary $\mathcal{V}$ with the same scaling dimension,
\begin{equation}
\mathcal{V} = (\partial_-\phi)^4 - \frac{5}{8\pi} \Big( (\partial_-^2 \phi)^2 - \frac{1}{5} \partial_-^2(\partial_-\phi)^2 \Big)
\end{equation}
From the thermal expectation values
\begin{equation}
\<T_{--}^2\> = \frac{27}{5} \varepsilon^2, \quad \<\mathcal{V}\> = 0,
\end{equation}
we can therefore derive (note that the thermal expectation value of any overall derivative is zero)
\begin{equation}
\<(\partial_-\phi)^4\> = 3 \varepsilon^2, \quad \<(\partial_-^2 \phi)^2\> = \frac{24\pi}{5} \varepsilon^2.
\end{equation}
In Figs.~\ref{fig_WeakVirasoro} (weak coupling) and \ref{fig_Virasoro} (strong coupling) we confirm that these expectation values scale as $\varepsilon^2$ and $\<\mathcal{V}\> \approx 0$ in high energy states. At lower temperatures, the deviation from $\langle \mathcal V\rangle = 0$ at weak coupling agrees with the analytic expectation for the free massive scalar. We expect that the similar deviation from $\langle \mathcal V\rangle = 0$ at strong coupling is also due to breaking of conformality, in this case by both dimensionful couplings $\lambda$ and $m^2$.

Comparing Figs.~\ref{fig_WeakVirasoro} and \ref{fig_Virasoro}, one also finds that the expectation values $\langle \mathcal{O}\rangle$ have a very weak dependence on the coupling $\lambda$ at high temperatures, in qualitative agreement with expectations given that the theory is asymptotically free.

The expecation value of $\phi^2$ is more subtle, as this is not a well-defined primary operator in the scalar CFT. However, because the Hamiltonian density $H \supset \frac{1}{2} m^2 \phi^2$, we can obtain its high temperature expectation value directly from the pressure
\begin{equation}
\<\phi^2\> = -2 \frac{\partial}{\partial m^2} P \approx \frac{T}{2m} \approx \sqrt{\frac{3\varepsilon}{2\pi m^2}}.
\end{equation}

\section{More on Eigenvector Statistics}\label{app_evecstatistics}

Exact diagonalization of the Hamiltonian provides not only the spectrum of eigenstates but also their wavefunction
\begin{equation}\label{eq_eigenstate}
|E\rangle = \sum_i | i \rangle c_{iE}\, , \qquad c_{iE} = \langle i | E \rangle\, ,
\end{equation}
in a computational basis $\{|i \rangle\}$ -- in the present approach this is the basis of primaries of the free scalar CFT. While the coefficients $c_{iE}$ evidently depend on the choice of this basis, one may expect eigenstates of chaotic systems to be approximately random in any reasonable computational basis. It is therefore interesting to compare statistical properties of the coefficients $c_{iE}$ to those of an eigenstate of a random matrix (in a fixed basis). This test of RMT is particularly appealing because it can be used to establish the thermality of an individual eigenstate. However, it is important to keep in mind that the Hamiltonian of a physical quantum many-body system has structure beyond that of a random matrix, which is reflected in its eigenvectors and eigenvalues. A simple example is the density of state, studied in Sec.~\ref{sec_thermo}, which must be taken into account (by unfolding the spectrum) in order to observe Wigner-Dyson statistics shown in Fig.~\ref{fig_WD}. More subtle deviations from RMT include the intermediate regime of hydrodynamics, discussed in Sec.~\ref{sec_discuss}, whereby spectral rigidity is suppressed for energy windows larger than the Thouless energy.

\begin{figure}
\centerline{
\begin{overpic}[width=0.9\linewidth,tics=10]{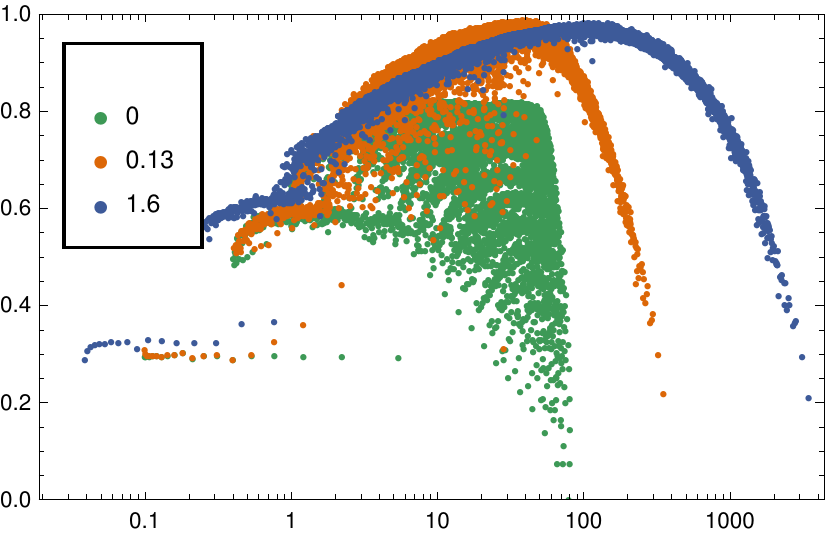}
	 \put (-5,25) {{\rotatebox{90}{$S_{|E\rangle}/S_{\rm GOE}$}}} 
	 \put (49,-2) {{$\varepsilon$}} 
	 \put (11,55) {{$\lambda/m^2$}} 
\end{overpic}}
\caption{\label{fig_evecstats_coeffentropy} Coefficient entropy of individual eigenstates normalized by the GOE value  (see Eq.~\eqref{eq_coeffentropy}) at zero, weak, and strong coupling ($\Delta=40$).}
\end{figure}

One simple probe of thermality of a given eigenstate \eqref{eq_eigenstate} is the coefficient entropy \cite{PhysRevE.81.036206,PhysRevE.85.036209}
\begin{equation}\label{eq_coeffentropy}
S_{|E\rangle}
	\equiv -\sum_i |c_{iE}|^2 \log |c_{iE}|^2\, .
\end{equation}
For chaotic physical systems, for the reasons described above, $S_{|E\rangle}$ is typically of the order of but smaller than its value for a GOE random matrix of the same size $N$, $S_{\rm GOE} = \log (0.48 N)$ \cite{PhysRevE.81.036206,PhysRevE.85.036209,DAlessio:2015qtq}. In Fig.~\ref{fig_evecstats_coeffentropy}, we show the coefficient entropy for eigenstates of the 1+1d QFT in Eq.~\eqref{eq_S}. At strong coupling $\lambda = 1.6 m^2$, one finds as expected that the coefficient entropy has low variance and is close but below the GOE value. At weak coupling, the variance increases and there are outliers (scar states), discussed in Sec.~\ref{sec_stats}. Finally, the free theory has a smaller coefficient entropy, with larger variance. We have found however that there are tests of chaos that more clearly distinguish the theory at various couplings, which we discuss next.

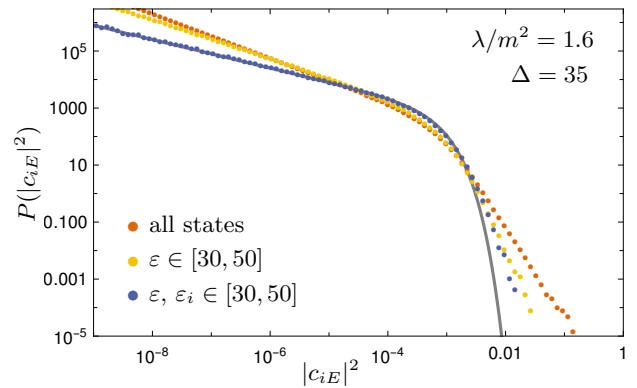
\begin{figure}
\centerline{
\begin{overpic}[width=0.9\linewidth,tics=10]{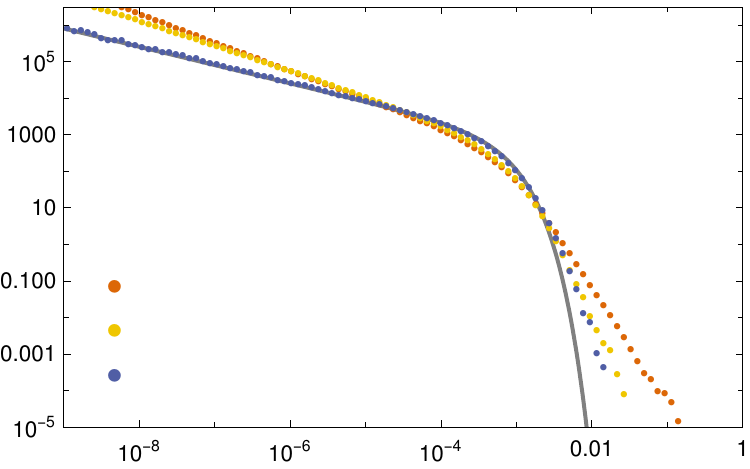}
	 \put (73,55) {{$\lambda/m^2 = 1.6$}} 
	 \put (79.8,49) {{$\Delta=35$}} 
	 \put (-5,25) {{\rotatebox{90}{$P(|c_{iE}|^2)$}}} 
	 \put (44,-2) {{$|c_{iE}|^2$}} 
	 \put (18,23.3) {{all states}} 
	 \put (18,17.5) {{$\varepsilon\in [30,50]$}} 
	 \put (18,11.6) {{$\varepsilon,\,\varepsilon_i\in [30,50]$}} 
\end{overpic}}
\caption{\label{fig_evecstats_app1} Distribution of eigenvector coefficients $|c_{iE}|^2 = |\langle i | E\rangle|$ at strong coupling. Performing the statistics over all coefficients (orange), one finds poor agreement with the GOE distribution (gray). The apparent polynomial tail at larger values of $|c_{iE}|^2$ was discussed in \cite{Srdinsek:2020bpq}. Restricting to eigenvectors $|E\rangle$ with energy density $\varepsilon = E/V$ in a small window partially suppresses this tail (yellow). If one further restricts the {\em basis} states to have average energy density $\varepsilon_i = \langle i| H | i \rangle/V$ in the same window, the tail is further suppressed and the distribution agrees well with GOE (blue). For these three sets of data one has $||P-P_{\rm GOE}||_1 = 0.55$ (orange), $0.36$ (yellow), and $0.052$ (blue).
}
\end{figure} 

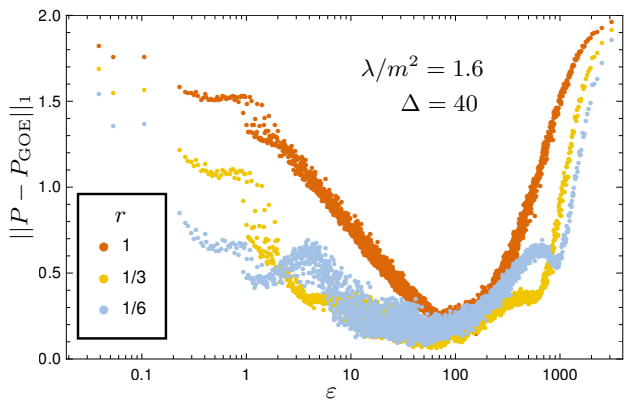
\begin{figure}
\centerline{
\begin{overpic}[width=0.9\linewidth,tics=10]{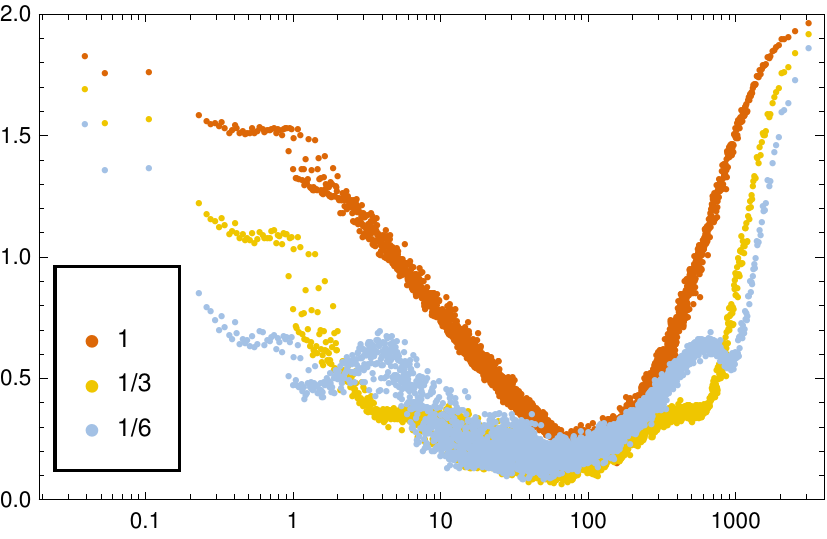}
	 \put (-5,25) {{\rotatebox{90}{$||P-P_{\rm GOE}||_1$}}} 
	 \put (49,-2) {{$\varepsilon$}} 
	 \put (13,28) {{$r$}} 
	 \put (55,53) {{$\lambda/m^2 = 1.6$}} 
	 \put (61.8,47) {{$\Delta=40$}} 
\end{overpic}}
\caption{\label{fig_evecstats_app2} Deviation of the coefficient statistics $P(|\langle E | i\rangle|^2)$ for individual eigenstates $|E\rangle$, from the GOE prediction. For each eigenvector $|E\rangle$, one performs the statistics over the coefficients $|\langle E|i\rangle|^2$, keeping only a ratio $r$ of all basis states $|i\rangle$ that have average energy $\langle i | H | i\rangle$ closest to $E$. Filtering basis states in this way (i.e. $r<1$) yields statistics that are more consistent with GOE. }
\end{figure}

As a more refined test of the thermality of an eigenstate $|E\rangle$, one can study the entire distribution of its coefficients $P(|c_{iE}|^2)$, and compare it to the distribution in RMT, Eq.~\eqref{eq:CoeffRMT}. In Ref.~\cite{Srdinsek:2020bpq}, it was found in several 1+1d QFTs that the distribution of coefficients did not agree with the GOE prediction, despite strong signatures of quantum chaos in other tests (such as eigenvalue spacing). However, as discussed above, the coefficient statistics of a QFT or any local quantum many-body system are not expected to converge to those of a random matrix, due to (among other things) a nontrivial equation of state and emergence of hydrodynamics at intermediate scales. It would be interesting to derive a theoretical expectation for the distribution $P(|c_{iE}|^2)$ that takes some of these effects into account, perhaps along the lines of what has been achieved with the spectral form factor \cite{Friedman:2019gyi,Winer:2020gdp} discussed in Sec.~\ref{sec_discuss}; or alternatively to establish a procedure to remove these effects and allow for direct comparison with RMT (similar to the unfolded eigenvalue spacing statistics \eqref{eq_unfolded}, which is insensitive to the density of states and hydrodynamics). In the absence of such a systematic procedure, we shall proceed heuristically as follows: given that RMT is only expected to describe the system in small energy windows (smaller than the Thouless energy), one should perform statistics only over correspondingly close eigenstates and basis states. Fig.~\ref{fig_evecstats_app1} shows that this intuition is largely correct. The coefficient statistics deviates significantly from the GOE expectation. However, if one restricts both the eigenstate $|E\rangle$ and basis state $|i\rangle$ to have (average) energy density in a fixed window, one finds remarkably good agreement with RMT. This motivates the following test of thermality of individual eigenstates: for each eigenstate $|E\rangle$, one studies the distribution of eigenvector coefficients $|c_{iE}|^2 = |\langle i| E\rangle|^2$ keeping only the fraction $0<r\leq 1$ of basis states $|i\rangle$ with average energy $\langle i |H|i \rangle$ closest to $E$. Fig.~\ref{fig_evecstats_app2} shows the result of this procedure, at different values of $r$. Especially at smaller energy densities $0.3\lesssim \varepsilon/m^2\lesssim 50$, one finds that eigenstates whose coefficient statistics strongly deviate from RMT in fact agree well with RMT if one restricts the basis states over which the statistics is performed. We interpret this result as indicating that a large window of energy densities (conservatively, $1\lesssim \varepsilon /m^2 \lesssim 100$) are chaotic -- a conclusion further supported by the fact that the ETH is satisfied in these states, as shown in Figs.~\ref{fig_collapse} and \ref{fig_collapse_app}. Furthermore, this test of thermality of individual eigenstates sharply distinguishes free, weakly coupled, and strongly interacting theories, as shown in Fig.~\ref{fig_evecstats}.

\vfill
\pagebreak
\bibliography{ETH_LCT}

\end{document}